\documentclass[sigconf]{acmart}
\usepackage{cleveref}

\usepackage{algorithm}
\usepackage{algpseudocode} 
\usepackage{booktabs}
\usepackage{makecell}
\usepackage{float}
\usepackage{booktabs}
\usepackage{multirow}
\usepackage{makecell}

\AtBeginDocument{%
  }

\copyrightyear{2026}
\acmYear{2026}
\setcopyright{cc}
\setcctype{by}
\acmConference[CHI '26]{Proceedings of the 2026 CHI Conference on Human Factors in Computing Systems}{April 13--17, 2026}{Barcelona, Spain}
\acmBooktitle{Proceedings of the 2026 CHI Conference on Human Factors in Computing Systems (CHI '26), April 13--17, 2026, Barcelona, Spain}
\acmPrice{}
\acmDOI{10.1145/3772318.3791844}
\acmISBN{979-8-4007-2278-3/2026/04}

\begin{document}

\title{InfoAlign: A Human–AI Co-Creation System for Storytelling with Infographics}

\author{Jielin Feng}
\email{jielinfeng23@m.fudan.edu.cn}
\orcid{0009-0008-1943-5609}
\affiliation{%
  \institution{School of Data Science, Fudan University}
  \city{Shanghai}
  \country{China}
}

\author{Xinwu Ye}
\email{xinwuye43@connect.hku.hk}
\orcid{0009-0008-3164-0373}
\affiliation{%
  \institution{School of Data Science, Fudan University}
  \city{Shanghai}
  \country{China}
}
\affiliation{%
  \institution{School of Computing and Data Science, The University of Hong Kong}
  \city{Hong Kong}
  \country{China}
}

\author{Qianhui Li}
\email{18812571619@163.com}
\orcid{0009-0006-8804-8130}
\affiliation{%
  \institution{School of Data Science, Fudan University}
  \city{Shanghai}
  \country{China}
}

\author{Verena Ingrid Prantl}
\email{verena.prantl@univie.ac.at}
\orcid{0009-0007-9792-729X}
\affiliation{%
  \institution{Faculty of Computer Science, Doctoral School Computer Science, University of Vienna}
  \city{Vienna}
  \country{Austria}
}

\author{Jun-Hsiang~Yao}
\email{rxyao24@m.fudan.edu.cn}
\orcid{0009-0000-8944-1942}
\affiliation{%
  \institution{School of Data Science, Fudan University}
  \city{Shanghai}
  \country{China}
}

\author{Yuheng Zhao}
\email{yuhengzhao@fudan.edu.cn}
\orcid{0000-0003-1573-8772}
\affiliation{%
  \institution{School of Data Science, Fudan University}
  \city{Shanghai}
  \country{China}
}

\author{Yun Wang}
\email{wangyun@microsoft.com}
\orcid{0000-0003-0468-4043}
\affiliation{%
  \institution{Microsoft Research Asia}
  \city{}
  \country{}
}

\author{Siming Chen}
\authornote{Siming Chen is the corresponding author.}
\email{simingchen@fudan.edu.cn}
\orcid{0000-0002-2690-3588}
\affiliation{%
  \institution{School of Data Science, Fudan University}
  \city{Shanghai}
  \country{China}
}
\affiliation{%
  \institution{Qingdao Research Institute, Fudan University}
  \city{Shanghai}
  \country{China}
}

\renewcommand{\shortauthors}{Feng et al.}
\renewcommand{\sectionautorefname}{Section}
\renewcommand{\subsectionautorefname}{Section}
\renewcommand{\subsubsectionautorefname}{Section}

\begin{abstract}
Storytelling infographics are a powerful medium for communicating data-driven stories through visual presentation. 
However, existing authoring tools lack support for maintaining story consistency and aligning with users' story goals throughout the design process. To address this gap, we conducted formative interviews and a quantitative analysis to identify design needs and common story-informed layout patterns in infographics. Based on these insights, we propose a narrative-centric workflow for infographic creation consisting of three phases: story construction, visual encoding, and spatial composition. Building on this workflow, we developed InfoAlign, a human–AI co-creation system that transforms long or unstructured text into stories, recommends semantically aligned visual designs, and generates layout blueprints. Users can intervene and refine the design at any stage, ensuring their intent is preserved and the infographic creation process remains transparent. Evaluations show that InfoAlign preserves story coherence across authoring stages and effectively supports human–AI co-creation for storytelling infographic design.

\end{abstract}


\begin{teaserfigure}
  \includegraphics[width=\textwidth]{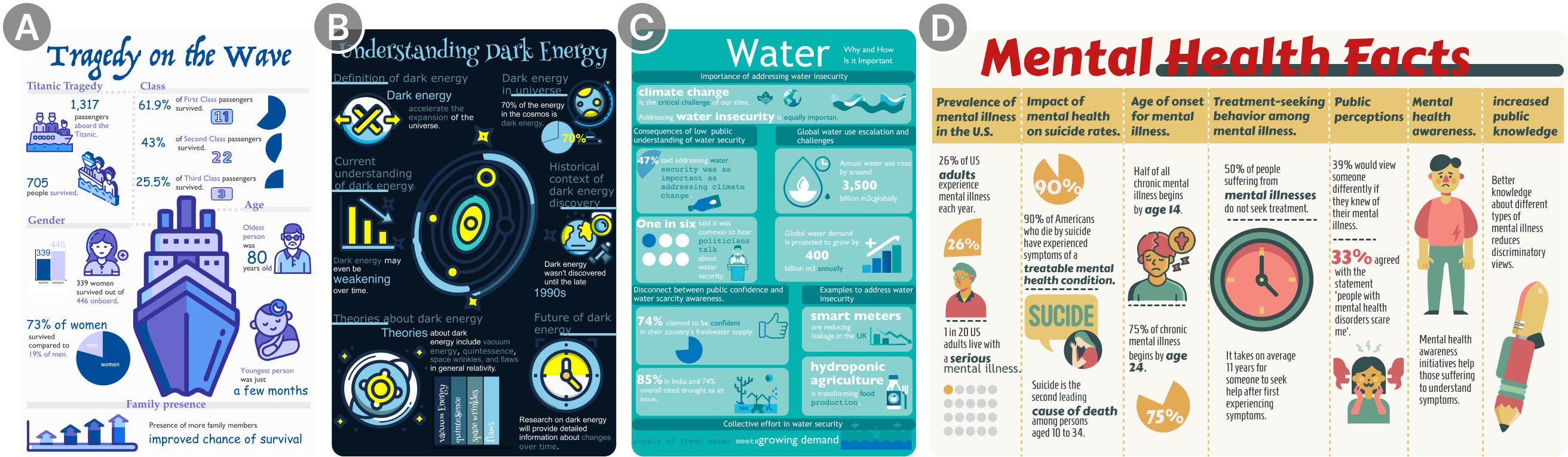}
  \caption{%
    Examples of storytelling infographics created by \textit{InfoAlign} from long or unstructured textual inputs and user queries across diverse domains:
(A) \textit{Tragedy on the Wave} (Titanic survival);
(B) \textit{Understanding Dark Energy} (astrophysical data);
(C) \textit{Water} (water usage);
(D) \textit{Mental Health Insights} (mental health key factors).
  }
  \Description{Examples of storytelling infographics created by InfoAlign from long or unstructured textual inputs and user queries across diverse domains. (A) Tragedy on the Wave tells the story of survival in the Titanic disaster through statistics. (B) Understanding Dark Energy unfolds a visual story that explains astrophysical data. (C) Water provide factual insights on water usage. (D) Mental Health Insights summarizes prevalence data and key factors related to mental health.}
  \label{fig:Case}
\end{teaserfigure}


\maketitle
\section{Introduction}

Infographics are visual representations designed to communicate complex information in a clear and engaging manner~\cite{harrison2015infographic}.
By integrating data, text, and visual design into a coherent structure, infographics are particularly effective at telling stories by transforming dense information into visual narratives that support specific communicative goals~\cite{marcus2015design,siricharoen2015infographic}. Such storytelling capability helps make information more accessible, memorable, and persuasive for broad audiences~\cite{lee2015more, lan2021smile, amit2018digital}, which has led to the widespread adoption of infographics in domains such as journalism~\cite{siricharoen2018infographic,de2018does}, education~\cite{ozdamli2018developing}, and advertising~\cite{andry2021interpreting}. 
\textcolor{black}{In practice, infographic creation unfolds through a multi-step, continuous design process in which design decisions are tightly interdependent. Design steps such as story construction, visual encoding, and layout arrangement are guided by a shared story goal, with earlier decisions shaping the direction and constraints of subsequent steps~\cite{zhou2024epigraphics, cao2023dataparticles}. At the same time, users refine and adjust their story intent as the design evolves, integrating it into successive design decisions. 
As a result, both design decisions and the user’s story intent must remain consistent across steps; otherwise, small misalignments can accumulate and fragment the story, ultimately undermining communication.
}

Given the opportunity, commercial infographic authoring tools such as
\textit{Piktochart}\footnote{\url{https://piktochart.com/}}, 
\textit{Venngage}\footnote{\url{https://venngage.com/}}, 
and \textit{Infogram}\footnote{\url{https://infogram.com/}} 
have lowered the technical barriers by automatically generating infographics from data. 
Similarly, recent research has formalized and automated specific infographic tasks, for example extracting salient facts from tabular data and composing them into fact sheets~\cite{wang2019datashot}, mapping proportion-related short textual statements to infographic visualizations~\cite{cui2019text}, transforming basic charts into embellished infographics~\cite{wang2018infonice, tyagi2022infographics}, and recommending visual assets or data filters based on a key message~\cite{zhou2024epigraphics}.
\textcolor{black}{However, infographic creation is inherently a continuous, multi-step process that requires story intent to be carried consistently across interdependent design decisions.
Commercial authoring tools largely rely on template-driven and highly automated approaches that often fail to support this consistency and offer limited opportunities for users to intervene during creation, while research systems tend to automate isolated infographic design stages based on specific tasks, providing limited support for maintaining story consistency across interdependent design stages.}

To tackle the challenge, the goal of our work is to propose an infographic generation system that enables the creation of infographics presenting consistent stories aligned with the user's story intent. 
Consequently, we address the following research questions.
\begin{itemize}
\item \textbf{RQ1} What design steps are necessary for an infographic to convey a story?
\item \textbf{RQ2} How can story consistency be maintained across the design steps of an infographic?
\item \textbf{RQ3} How can the user's story intent be preserved throughout the design steps of an infographic?
\end{itemize}

To investigate \textbf{RQ1}, we conducted formative interviews with eight professional infographic designers and examined their workflows and challenges. The design process typically began with story structuring, where designers struggled to select relevant information from large or fragmented text and organize it into a story. The next step was visual design, as designers often found it difficult to create elements that align with the stories. Finally, during layout arrangement, they encountered difficulties composing these designs into layouts that convey a clear and consistent story flow.
Based on the identified design steps, we further investigated \textbf{RQ2}. We conducted a quantitative analysis of \textcolor{black}{70} real-world storytelling infographics to extract common structural and layout patterns. Building on these patterns, we propose a narrative-centric workflow that structures infographic creation into three phases. The process consists of story construction, visual encoding, and spatial composition, ensuring that stories are consistently preserved across design steps. It first organizes long or fragmented text into a coherent story, then generates visual designs aligned with the story, and finally arranges them into layouts that reflect the story flow. Building on this narrative-centric workflow, we developed InfoAlign, a human–AI co-creation system for the creation of infographics that tell stories. InfoAlign addresses \textbf{RQ3} by preserving the user's story intent through a combination of AI recommendations and human interventions. Users can adjust the constructed story, refine visual designs, and choose layouts that match their narrative goals, with the final output remaining editable on a free canvas. We conducted a user study with 12 participants to (1) demonstrate InfoAlign's workflow-level quality in maintaining coherent stories across design process, (2) examine its human–AI co-creation patterns, and (3) validate its overall effectiveness in supporting the creation of infographics that present consistent stories aligned with users’ story intent. The key contributions of this paper are:
\begin{itemize}
\item \textbf{A narrative-centric workflow} for storytelling infographic creation that preserves consistent stories across design steps, consisting of three core phases: narrative construction, visual encoding, and spatial composition.

\item \textbf{InfoAlign}, a human–AI co-creation system that transforms textual input and user queries into storytelling infographics and preserves users' story intent through human-in-the-loop intervention at each step.

\item \textcolor{black}{\textbf{A user study} that demonstrates InfoAlign's workflow-level quality in maintaining consistent stories throughout the design process, reveals human–AI interaction patterns during creation, and evaluates the system's overall effectiveness.}
\end{itemize}

\section{Related Work}\label{sec-2}
This section reviews three important areas of research that inform our work: Generative AI for Visualization, Infographic Design, and Narrative Visualization.

\subsection{Generative AI for Visualization}
Recent advancements in artificial intelligence (AI) have fundamentally advanced both text and image generation. Large language models (LLMs) such as GPT-4 \cite{bubeck2023sparks} and GPT-4o \cite{openai2024gpt4o} demonstrate strong capabilities in text generation and reasoning. In parallel, progress in image generation, particularly diffusion-based methods \cite{nichol2021improved,song2020denoising}, has enabled controllable and expressive visual outputs. For example, \textit{DALL-E 3} \cite{openai2023dalle3} produces diverse, high-quality outputs, and \textit{ControlNet} \cite{zhang2023adding} provides superior control over the generation process. Unlike conventional pixel-based models, our work leverages \textit{RecraftAI} \cite{recraftai} to generate scalable vector graphics, which are well suited for infographic design due to their editability and clarity. Together, these advances provide a foundation for AI-assisted infographic creation.

Numerous studies~\cite{zhou2024stylefactory,brade2023promptify,ye2024generative,wu2021ai4vis} have explored applying generative AI to stylized visualization and creative design. For instance, \textit{Let the Chart Spark} \cite{xiao2023let} generates pictorial visualizations by aligning semantic context with data, and \textit{Viz2Viz} \cite{wu2023viz2viz} transforms existing charts through text-guided cues to support flexible re-styling. Other systems extend generative methods to specific domains: \textit{LIDA} \cite{dibia2023lida} automates data-driven infographic generation, \textit{PlantoGraphy} \cite{huang2024plantography} integrates iterative interaction to configure generative models for human-centered design, and \textit{TypeDance} \cite{xiao2024typedance} 
creates semantic typographic logos through personalized image-guided generation. 

\textcolor{black}{Previous LLM- or VLM-assisted visualization systems typically treat generative models as end-to-end visual creators. In contrast, InfoAlign adopts a hybrid workflow that combines LLMs for information extraction and visual asset generation with layout algorithms for global structure, mitigating the tendency of generative models to produce layouts that mismatch the intended story pattern. Building on this hybrid workflow, the InfoAlign system supports human–AI co-creation: models offer recommendations and provide guidance, while users intervene and iterate as equal partners to keep the infographic aligned with their story intent.}

\subsection{Infographic Design}
Infographics are structured visual representations that integrate data visualizations, text, and design elements to communicate complex information engagingly \cite{lan2021smile,andry2021interpreting,nuhouglu2024infographic}. By breaking down intricate topics into accessible visuals, they play a vital role in conveying data-driven narratives \cite{marcus2015design,siricharoen2015infographic}. Prior work highlights their benefits, including improved long-term recall \cite{bateman2010useful}, reduced comprehension barriers \cite{marcus2015design}, and increased audience engagement \cite{andry2021interpreting}.

Despite these benefits, creating infographics remains challenging due to their inherently multi-step and complex design process. Researchers have explored automated approaches inspired by templates and design practices. For instance, \citet{zhu2019towards} extracted extensible timeline templates from bitmap images, \textit{Vistylist} \cite{shi2022supporting} applied style transfer to pictorial visualizations while preserving content integrity, and \citet{qian2020retrieve} generated proportion-based infographics by retrieving and adapting online examples. Existing research has also produced various tools for direct infographic generation. 
For example, \textit{Epigraphics} \cite{zhou2024epigraphics} leverages text-based instructions to guide creative infographic design. 
\textit{Text-to-Viz} \cite{cui2019text} generates proportion-related infographics from short textual statements, 
\textit{InfoNice} \cite{wang2018infonice} enables the creation of data-driven information graphics from basic charts, 
and \textit{Datashot} \cite{wang2019datashot} produces fact sheets from tabular input. 
Beyond infographic generation, some studies focus on layout composition. 
For instance, \citet{lu2020exploring} examined visual information flow in infographics, 
and \textit{Infographics Wizard} \cite{tyagi2022infographics} automatically recommends infographic layouts.



\textcolor{black}{While prior infographic automation tools have generated infographics from inputs such as tabular data or basic charts, they have largely remained task-based, emphasizing statistical representation or visual expressiveness rather than supporting data-driven storytelling. This gap reflects the absence of mechanisms for modeling story goals or maintaining narrative coherence throughout the design process. In contrast, InfoAlign extends this line of work with a storytelling-oriented approach. We take long and unstructured text, together with a user-specified story goal, as input and introduce a narrative-centric workflow that consistently organizes and expresses the information contained in these inputs as a coherent visual story across the entire creation process.}

\subsection{Narrative Visualization}
Narrative visualization combines data and storytelling to communicate intended messages effectively \cite{sun2022erato,segel2010narrative,lee2015more}. Recent work emphasizes integrating storytelling with visualization to enhance user engagement and comprehension \cite{kosara2013storytelling}. Several theoretical frameworks have been proposed to support visual data story creation. For instance, \citet{chen2023does,li2024we} examine how human–AI collaboration can facilitate narrative visualization, while \citet{outa2024highlighting} introduces a structured process for narrative construction. Other studies identify low-level analysis tasks for developing stories \cite{amar2005low,chen2009toward}, and reviews of natural language generation highlight its role in enriching narrative content \cite{hoque2025natural,shen2022towards}.

Given the complexity of integrating data, visualization, and narrative presentation \cite{li2024we,chevalier2018analysis}, a variety of tools have been developed. For example, \textit{LightVA} \cite{zhao2024lightva} provides a framework for task proposal and story modeling, while \textit{Calliope} \cite{shi2020calliope} automatically generates series visulizations from spreadsheet inputs, and \textit{Erato} \cite{sun2022erato} supports interactive story editing and fact interpolation. Other systems, such as \textit{DataNarrative} \cite{islam2024datanarrative} and \textit{FinFlier} \cite{hao2024finflier}, employ multi-agent frameworks for domain-specific narratives. Tools like \textit{Charagraphs} \cite{masson2023charagraph} and \textit{Notable} \cite{li2023notable} provide real-time narrative assistance. Beyond these, alternative presentation formats have also been explored, including automated sports news generation \cite{cheng2024snil,liu2022opal}, data-driven videos \cite{liao2022realitytalk,shao2025narrative,wang2024wonderflow,zheng2022telling}, and comic-style visual narratives \cite{zhao2021chartstory}.


Although existing studies have advanced narrative visualizations, \textcolor{black}{they largely concentrate on sequences of visualizations and charts. Compared with visualization-authoring tools that generate individual narrative views, our work treats storytelling infographics as the outcome of a multi-step creation process in which user intent must be preserved across story construction, visual design, and spatial composition. InfoAlign supports this process by allowing users to intervene at each step, so that decisions throughout the design process can be continuously grounded in and refined by their own  story intent.}

\section{Formative Interview}
\label{sec-3}
The creation of infographics that present consistent stories remains challenging and resource-intensive. To gain deeper insights into the requirements and obstacles involved, we conducted formative interviews with eight infographic creators, aiming to understand current practices and explore how Human-AI collaboration could facilitate the creation of effective storytelling infographic.

\textbf{Participants}
The formative interview involved eight infographic creators (labeled P1 to P8), including two males (P1, P7) and six females (P2, P3, P4, P5, P6, P8). Participants' ages ranged from 21 to 39 years ($M = 28.25, SD = 6.93$).
Six participants (P1–P6) were industry professionals who designed infographics or posters for industrial purposes. Their experience varied: P1 had 10 years of experience and designed infographics several times a year, while P4, also with 10 years of experience, designed infographics several times a month. P5 (5 years of experience) and P6 (3 years of experience) also reported creating infographics several times a month. P2 and P3, both with 2 years of industry experience, created infographics several times every 3 months. The remaining two participants (P7, P8) were affiliated with universities, where their infographic creation supported academic competitions or research activities. Both had 2 years of experience: P7 designed infographics several times a month, whereas P8 created them several times per season.

\textbf{Procedure}
To explore their creation processes, requirements, and challenges, we conducted semi-structured interviews with each participant using a pre-defined set of questions. We asked about (Q1) the types of input data used, (Q2) their general workflow during infographic creation, (Q3) the key components they considered crucial for design, and (Q4) their perspectives on what defines a ``good'' infographic. We also gathered feedback on (Q5) the AI-powered and non-AI-powered tools participants had used for creating or assisting infographic design. Additionally, we asked (Q6) about the challenges they encountered and (Q7) their expectations for the functionality, interaction design, and user interface of an AI-powered infographic authoring tool. Each participant answered all question, and interviews were video-recorded for later analysis.

\subsection{Findings}
\label{sec-3.1}
Based on interviews with participants P1–P8, we summarize the key findings \textcolor{black}{(F1-F5)} as follows.

\textbf{\textcolor{black}{F1}: Conceptualizing Infographic Design as Storytelling Processing.}
Our findings suggest that infographic design can be conceptualized as visual storytelling, with creators focusing on shaping and conveying a story. They typically begin by defining a clear \textbf{goal} (Q2), which serves as the story goal of the infographic. Except for P3, all participants emphasized the importance of articulating this goal early on as the foundation of their work. Once the goal is set, participants generally move through three interconnected stpes: \textit{Story Structuring}, \textit{Visual Design}, and \textit{Layout Arrangement} (Q2). Story structuring involves constructing the story based on the goal, selecting salient content, and organizing it into a coherent narrative. Visual design focuses on transforming the story into visual forms by choosing thematic colors and fonts, designing highlights that emphasize key messages, creating charts to convey insights, and using icons or graphics that reflect the story's semantic (Q3). Layout arrangement brings all designs together by arranging text, visuals, and charts into a layout that reflect the story flow. P7 illustrated this process: ``I first clarify the goal of the infographic. Based on this goal, I process the data into a story structure that conveys my intended message. I then present insights with text and charts and select design elements such as colors, illustrations, and icons that capture the story semantics. Lastly, I arrange all designs into a layout that reflects the story flow.''

Participants further noted that the design process (Q2) and the key components (Q3) directly shaped their view of what makes a good infographic (Q4). As P1 emphasized, a infographic must first ensure that its story is clearly conveyed and aligned with the intent goal; only then do aesthetics and readability come into play. P5 echoed this idea: ``Infographic: first `Info,' then `Graphic.' A clear story comes first; visual expressive follows.''

\textcolor{black}{\textbf{F2: Complex Textual Inputs and the Challenges of Constructing Stories.}}
Our findings indicate that textual data are one of the primary input formats utilized in storytelling infographic design (Q1). Specifically, P1, P3, P5, P7 reported experience with both textual and tabular data inputs, P2, P4, P8 exclusively worked with textual data, and P6 only used tabular data. P5 explained, ``I primarily use textual data provided by other departments; tabular data is rarely used.''
While previous research has frequently highlighted the use of tabular data as a common input for infographic creation\cite{wang2019datashot,zhou2024epigraphics}, our results revealed that complex textual data input plays a significant role in the storytelling infographic creation process. \textcolor{black}{Most participants (P1, P2, P3, P4, P5, P8) emphasized that textual materials were central to storytelling infographics. As P2 noted, ``Many real-world infographic projects began with narrative reports, online articles, or mixed-format briefs.''} We identified two key characteristics of such input: \textbf{\textit{long}} and \textbf{\textit{unstructured}}. Long text refers to extensive documents, such as multi-page Word or PDF files containing rich information. Unstructured text refers to fragmented pieces of information related to a common theme or topic but lacking a clear organizational order. As P1 described, ``I copied information from several Wikipedia pages and compiled it into a PDF file as raw data for the infographic.'' 

\textcolor{black}{Participants also reported using a variety of tools, both AI-powered and non-AI-powered, to support infographic creation (Q5). Commonly used non-AI tools included D3.js, Adobe Illustrator, Adobe Photoshop, Canva, and Figma, along with online platforms offering templates and asset libraries. P4 and P5 also described their experiences with AI-powered tools, such as \textit{Piktochart} for simple infographic generation, Gaoding Design for AI-generated templates, and the AI-assisted features in Canva for creating graphic assets. However, despite the availability of these tools, participants emphasized persistent challenges, particularly when working with long or unstructured data. Participants noted that extracting relevant information and shaping it into a story requires considerable effort (Q6). As P1 explained, ``It is hard for me to select and extract the key information to clearly illustrate my story goal.'' Similarly, P8 highlighted the difficulty of handling large volumes of unstructured data and organizing them into a meaningful story.}

\textbf{\textcolor{black}{F3}: The Need for Visual Designs Aligned with Story and User Intent.}
For visual designs, participants reported persistent challenges in aligning their choices with both the unfolding story and their intended goals (Q6). They struggled to select elements such as colors and fonts, and to design components such as backgrounds, highlights, charts, and icons in ways that meaningfully supported each part of the story. P2 expressed frustration: ``I always struggle with choosing suitable colors or graphical components for different pieces of information in the story. It is often complicated and time-consuming, and the visuals usually fail to match the specific themes of each part of the story.''

\textbf{\textcolor{black}{F4}: Expectations for Layout Recommendation and Automation that Support Story Reading Flow.}
Regarding layout arrangement, participants emphasized the need for layouts that guide readers through the story with a clear and natural reading flow (Q6). P3 noted: ``Layout arrangement is the most difficult part for me. I often struggle to position components so that they not only look good but also guide readers through the story in a clear way. I really wish layouts could be generated automatically with some guidance, instead of relying on rigid templates that feel unnatural and disconnected.''

\textbf{\textcolor{black}{F5}: Expectations for Human-AI Collaboration in Infographic Creation.}
Participants envisioned a need for human–AI collaboration in infographic creation (Q7). They expected AI to automatically structure stories from input data, suggest and generate visual designs, and arrange these designs into layouts. At the same time, users emphasized the importance of being able to intervene at every step: adjusting story structure and content, refining visual designs to express their individual intent, and selecting or previewing layouts. As P5 summarized: ``AI should quickly generate an storytelling infographic from raw data, while allowing me to refine and personalize the design at each step. Furthermore, I expect the entire creation process, from data upload to the final infographic output, to be seamlessly integrated within a single platform.''

\subsection{Design Requirement}
\label{sec-3.2}
Drawing from these findings, we derived five design requirements (R1–R5) for the storytelling infographic design process.

\textbf{R1: Support Long or Unstructured Textual Data with User Queries as Input.}
\textcolor{black}{Since participants relied mainly on textual data, the system should primarily} support long and unstructured textual input combined with user queries, \textcolor{black}{while tabular data plays an auxiliary role and can be converted into text when necessary.}
(\textbf{\textcolor{black}{F2}}).  


\textbf{R2: Enable Coherent Story Structuring.}  
The system should help users transform long or unstructured data into a coherent story structure, ensuring that key points are selected, organized, and clearly expressed (\textbf{\textcolor{black}{F1, F2}}).  


\textbf{R3: Provide Visual Designs Aligned with Story and User Intent.}  
The system should recommend visual designs (e.g., colors, fonts, highlights, icons, and charts) that align with both the story and user intent, ensuring thematic consistency (\textbf{\textcolor{black}{F1, F3}}).  


\textbf{R4: Recommend Layouts that Guide Reading Flow.}  
The system should generate layouts that are not only visually natural but also support a logical reading flow, helping viewers follow the story while avoiding rigid, overly templated designs (\textbf{\textcolor{black}{F1, F4}}).  


\textbf{R5: Support Seamless Human–AI Collaboration with Flexible Control.}  
The system should allow AI to automate story structuring, visual design, and layout generation, while enabling users to intervene and refine at each step, ensuring that the final infographic reflects their own story intentions (\textbf{\textcolor{black}{F5}}). 


\section{Layout Design-Space Analysis}
\label{sec-4}
To address the design requirements in \cref{sec-3}, storytelling infographics need a clear story structure \textbf{(R2)}, visual designs aligned with the story and user intent \textbf{(R3)}, and layouts \textbf{(R4)} that guide reading flow. Prior work has offered partial guidance on visual designs for storytelling \cite{zhou2024epigraphics} \textbf{(R3)}, and has explored component arrangement and general layout structures \cite{cui2019text, lu2020exploring}, but overlooked how layouts convey story flow \textbf{(R2, R4)}. To understand how story structures and layouts support story flow in infographics, we analyzed \textcolor{black}{70} real-world examples and derived structural and layout patterns for visual storytelling.

\begin{figure}[h]
    \centering
    \includegraphics[width=\linewidth]{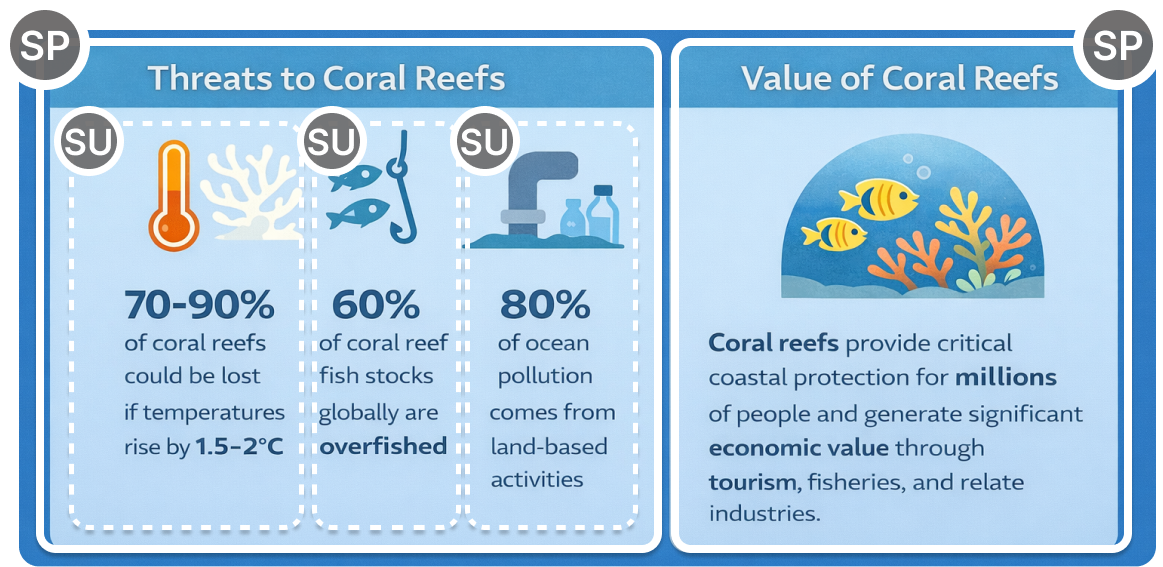}
    \caption{%
    Illustrative example of breaking storytelling infographic cases into Story Piece (SP) and Story Unit (SU).
    }
    \label{fig:StudyExample}
    \Description{Illustrative example of breaking storytelling infographic cases into Story Piece (SP) and Story Unit (SU).}
\end{figure}

\subsection{Data Preparation and Coding Process}
\label{sec-4.1}
We clarify our coding process by \textcolor{black}{outlining the data collection and filtering steps, key term definitions, coding procedure, class imbalance handling, and the final dataset.}

\subsubsection{Data Collection and Filtering}
\label{sec-4.1.1}
\ 
\newline
\indent 
\textbf{Data Source and Initial Collection.} Since infographics are widely used in practice, many design examples can be found online. We chose Pinterest as our source as it hosts a large collection of creative, real-world infographics. \textcolor{black}{and prior work on creative visualizations has followed similar sourcing practices \cite{ying2022metaglyph}.} Using the keyword ``infographic'', we collected 200 examples.

\textbf{Inclusion and Quality Criteria.}
To ensure the collected examples satisfy creators' storytelling infographic requirements (\cref{sec-3.1}), we derived inclusion criteria from the design requirements in \cref{sec-3.2}. Specifically: 1) The infographic must present a clear story goal (R1). 2) The infographic should be plausibly derivable from textual data, based on what can be inferred from its visual content (R1). 3) The infographic must present several related key points that are visually separated into sections, all contributing to the same coherent story (R2). 4) The infographic must use visual elements such as colors and icons that reinforce the story (R3). 5) The infographic must adopt a layout with a clear narrative sequence and reading flow, while avoiding rigid or overly templated designs (R4). In addition, to ensure example's quality, 6) the infographics must be high-resolution with clearly readable text, complete visual composition, and clear evidence of intentional infographic design rather than templated, or low-effort works.



\textbf{Coder Training and Data Filtering.}
Two trained coders (C1 and C2), with 7 and 6 years of infographic experience respectively, participated in both the filtering stage and the subsequent segmentation and layout coding. Before filtering, they were trained on the inclusion criteria and calibrated on several pilot cases. Each coder then independently identified which of the 200 collected infographics should be included, yielding an inter-rater agreement of Cohen's $\kappa = 0.68$. Disagreements were resolved through discussion, resulting in 48 examples that clearly met our definition of storytelling infographics \textcolor{black}{(See Sec. E1 in the supplementary material for the detailed codebook and examples.)}

\subsubsection{Term Definitions}
\label{sec-4.1.2}
\ 
\newline
\indent 
We defined the concepts of \textbf{Storytelling Infographic}, \textbf{Story Piece (SP)}, \textbf{Story Unit (SU)}, \textbf{Visual Design (VD)}, \textbf{Narrative Logic}, \textbf{Data Insight}, and \textbf{Storytelling Infographic Layout}, and used these definitions as the basis for our coding process.

\textbf{Storytelling Infographic}:
A storytelling infographic consists of multiple \textit{Story Pieces} that together form the overall story framework.

\textbf{Story Piece (SP)}:
Story pieces are narratively related to the user's story goal for the infographic based on the \textit{Narrative Logic}.

\textbf{Story Unit (SU)}:
Story Units each represent a distinct \textit{data insight} derived from their corresponding \textit{Story Piece}. A Story Unit is composed of multiple \textit{Visual Designs}.

\textbf{Visual Design (VD)}:
Visual Designs include \textit{Highlight Text}, \textit{Regular Text}, \textit{Icons/Graphics}, and \textit{Charts}.

\textbf{Narrative Logic} refers to the narrative relationship among Story Pieces. We reviewed commonly used coherence relations as defined in discourse studies \cite{wellner2006classification, wolf2005representing, shi2020calliope}, and employed nine types of narrative logic commonly observed in the collected storytelling infographics, \textcolor{black}{namely Silimarity, Cause-effect, Contrast, Violated Expectation, Temporal, Attribution, Example, and Generalization (See Sec. E1 in the supplementary material for detailed term definitions.)}

\textbf{Data Insight} refers to the type of insight contained within each Story Unit. We investigated and adopted nine types of data insights based on prior studies \cite{amar2005low, chen2009toward, wang2019datashot}, \textcolor{black}{including Value, Difference, Proportion, Trend, Categorization, Distribution, Rank, Extreme, and Textual Statement. (See Sec. D2 in the supplementary material for detailed term definitions.)}

\begin{figure*}[t]
    \centering
    \includegraphics[width=\linewidth]{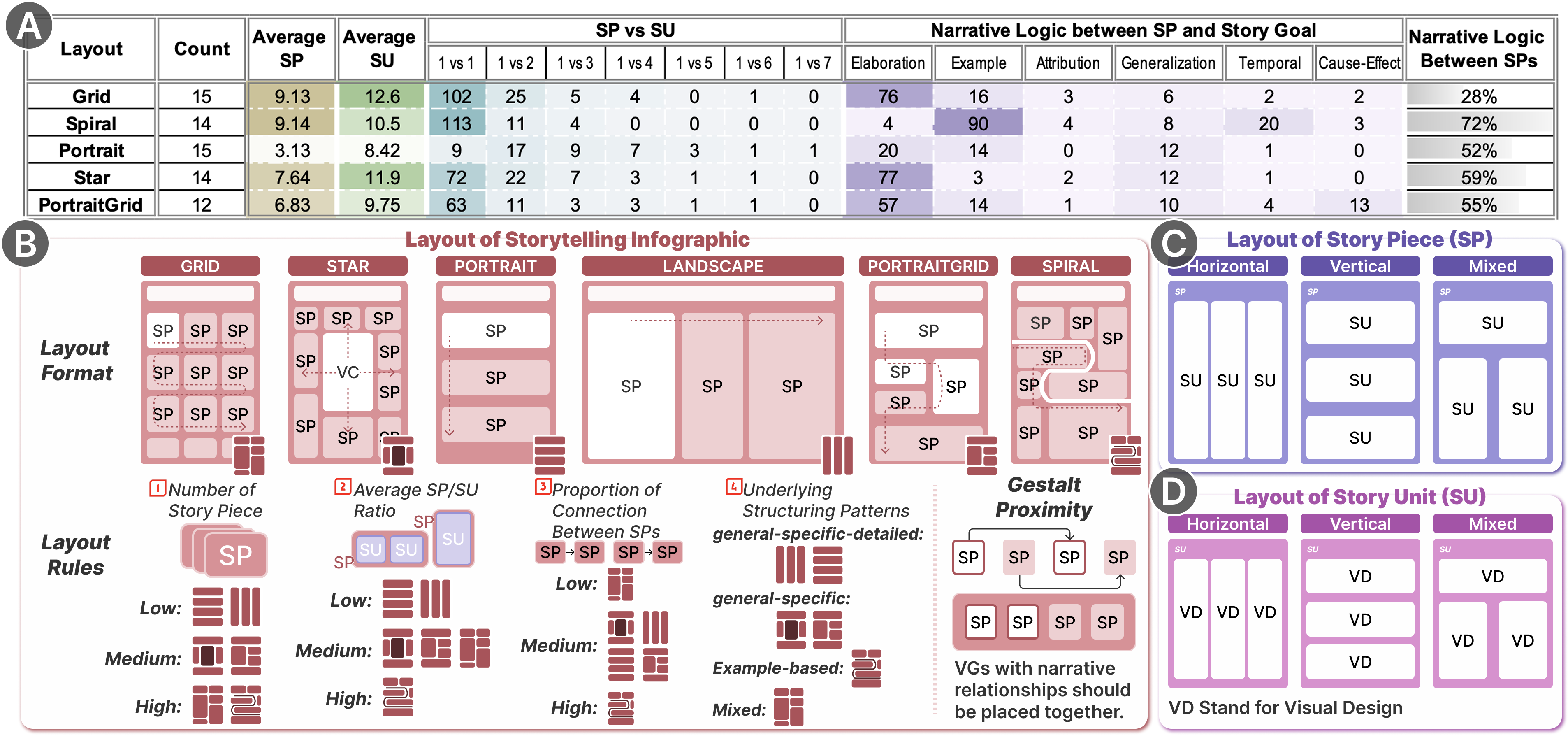}
    \caption{%
    Overview of the quantitative analysis and layout patterns observed in storytelling infographics.
    (A) Aggregated results comparing average numbers of Story Pieces (SPs) and Story Units (SUs), as well as proportions of narrative logic under different layout types.
    (B) Illustration of six storytelling infographic layout (\textit{Grid}, \textit{Star}, \textit{Portrait}, \textit{Landscape}, \textit{Spiral}), and \textit{PortraitGrid}), along with key design rules.
    (C) Configurations for arranging SUs 
    (D) Configurations for arranging VDs (icons, text, charts, and highlights.)
    }
    \label{fig:Analysis}
    \Description{Overview of the quantitative analysis and layout patterns observed in storytelling infographics. (A) Aggregated results comparing average numbers of Story Pieces (SPs) and Story Units (SUs), as well as proportions of narrative logic under different layout types. (B) Illustration of six storytelling infographic layout (Grid, Star, Portrait, Landscape, Spiral), and PortraitGrid), along with key design rules. (C) Configurations for arranging SUs (D) Configurations for arranging VDs (icons, text, charts, and highlights.)}
\end{figure*}

\textbf{Storytelling Infographic Layout}: 
Since storytelling infographics convey narratives through composite Story Pieces (SPs), the arrangement of SPs is what primarily determines the reader's overall story flow. We therefore do not analyze layout patterns within SUs or VDs. Instead, we adopt standard horizontal, vertical, and mixed arrangements for VDs within SUs and for SUs within SPs~\cite{cui2019text} (see \cref{fig:Analysis}(C), (D)). Our primary focus is on shaping the reader's story flow by composing SPs into a coherent story structure.
Following the categorization proposed by \cite{lu2020exploring}, we investigated and employed six types of  storytelling infographic layouts. \textbf{\textit{Grid Layout}} refers to a layout where each row or column contains two or more grid cells, with each cell containing one SP. The reading order typically follows a top-left to bottom-right pattern. \textbf{\textit{Star Layout}} refers to a layout with a central visual element (usually a graphic) surrounded by SPs, each semantically linked to the center so that the story radiates outward from this core. \textbf{\textit{Portrait/Landscape Layout}} refers to a layout in which each row or column contains only one SP, forming a vertical (portrait) or horizontal (landscape) sequence. The reading order typically proceeds from top to bottom or left to right. \textbf{\textit{PortraitGrid Layout}} refers to a hybrid layout that combines features of Portrait and Grid layouts. Some rows contain a single SP, while others contain multiple SPs, often with single-SP rows at the top or bottom and grid-like rows in the middle. The reading order usually goes from top to bottom and, within each row, from left to right. \textbf{\textit{Spiral Layout}} refers to a layout guided by a curved path, with SPs placed sequentially along the spiral. The story flow follows the curve from its starting point to its endpoint.

\subsubsection{Coding Procedure}
\ 
\newline
\indent 
\textcolor{black}{For the 48 collected infographics, we segmented and annotated each example according to the term definitions, coded their layout patterns, and refined the dataset by supplementing under-represented layout types.}

\textbf{Infographic Segmentation and Annotation.}
We segmented every examples into Story Pieces (SPs) based on Narrative Logic, and further into Story Units (SUs) based on Data Insight (\cref{fig:StudyExample}). \textcolor{black}{SP/SU segmentation involves interpretive boundary decisions grounded in narrative logic and data insight definitions rather than simple category assignment. We therefore adopted a consensus coding procedure commonly used in structural and qualitative analysis \cite{macqueen1998codebook}. The first coder (C1) performed the initial segmentation, and annotating the narrative logic relation between each SP and the overall story goal and among SPs. The second coder (C2) independently reviewed all segmentations and annotations, and any disagreements, typically concerning fine grained boundary placement or the choice among closely related narrative logic relations, were resolved through discussion until full consensus was reached} \textcolor{black}{(See Sec. E2 in the supplementary material for the detailed codebook and examples.)}

\textbf{Infographic Layout Coding.}
\textcolor{black}{C1 and C2 then independently coded the layouts of 48 infographic examples (Cohen's $\kappa = 0.92$). The few disagreements were resolved through discussion until consensus was reached.}

\textbf{Purposive Keyword-Guided Sampling for Under-Represented Layouts.}
After coding the initial 48 infographic examples (15 Grid, 2 Spiral, 5 Portrait, 14 Star, 12 PortraitGrid), we found that spiral and portrait layouts were substantially under-represented. Although this imbalance may reflect their lower prevalence online, our goal is to analyze how different layouts support storytelling rather than to estimate natural layout distributions; thus, relying on only a few cases would risk unstable pattern identification. To enrich these two categories, C1 and C2 independently conducted purposive, keyword-guided searches \cite{patton2014qualitative} \textcolor{black}{on Pinterest using terms such as ``infographic with curved flow'' and ``top-to-down infographic.''}
\textcolor{black}{These examples were not readily retrievable via the generic ``infographic'' keyword alone, likely due to coarse tagging practices and platform-specific ranking algorithms.}
After deduplication, 68 unique candidates remained. Applying the same filtering procedure described in \cref{sec-4.1.1} yielded 24 valid examples (Cohen's $\kappa = 0.76$). We then applied the same segmentation and annotation procedure. Based on the structural definitions of spiral and portrait layouts, two candidates were excluded for not meeting the required SP arrangement patterns, resulting in 22 additional examples (10 portrait, 12 spiral) added to the final dataset.

\subsubsection{Final Dataset and Analysis Measures}
\ 
\newline
\indent 
As shown in \cref{fig:Analysis}, \textcolor{black}{our final storytelling infographic dataset contains 70 infographic examples.} We calculated the average number of Story Pieces (Average SP) and the average number of Story Units (Average SU) per infographic. Additionally, we documented the number of Story Units contained within each Story Piece across all examples (SP vs. SU).
We also recorded the Narrative Logic between each Story Piece and the overall Story goal (Narrative Logic between SP and Story Goal). Furthermore, we recorded the narrative connections between Story Pieces. Specifically, for each infographic, we computed the proportion of Story Pieces that exhibited a narrative relation with one other Story Piece. We then averaged these proportions across all infographics within the same layout type to obtain the proportion of Narrative Logic between SPs (Narrative Logic between SPs).

\subsection{Layout Patterns for Storytelling Infographics}
\label{sec-4.2}
We identified a set of infographic layout patterns grounded in our coding results, which later informed the layout recommendation rules for InfoAlign. These patterns take into account the number of story pieces, the number of story units within each piece, the proportion of narrative connections across pieces, the underlying structuring patterns, and the Gestalt principle of proximity (\cref{fig:Analysis}(B), Layout Rules and Gestalt Proximity).

\textbf{The Number of Story Pieces.}
Our findings show that the number of story pieces affects the layout of storytelling infographics. Since each row or column accommodates only one story piece, the \textbf{\textit{Portrait}} and \textbf{\textit{Landscape}} layouts are more suitable for stories with fewer story pieces (\textcolor{black}{$M = 3.13$}). For stories with a moderate number of story pieces, the \textbf{\textit{Star}} and \textbf{\textit{PortraitGrid}} layouts are more appropriate ($M = 7.64$ and $M = 6.83$, respectively). The Star layout uses a central visual graphic to anchor the design, while the PortraitGrid layout arranges pieces across multiple columns, offering flexible space distribution. When dealing with a larger number of story pieces, the \textbf{\textit{Grid}} and \textbf{\textit{Spiral}} layouts are more effective ($M = 9.13$ and \textcolor{black}{$M = 9.14$}, respectively). Their flexible structures allow story pieces to be placed side-by-side in rows or along curves. Additionally, the layout category with the highest average number of story pieces supports approximately \textcolor{black}{9.14} SPs. Therefore, in InfoAlign, we set an upper limit of 10 story pieces to ensure readability.

\textbf{The Density of Story Units Within Story Pieces.}
Our findings reveal that the density of story units within each story piece (i.e., the ratio between story pieces and story units) influences the choice of layout. A lower average ratio indicates that more story units are contained within each story piece, while a higher ratio suggests fewer units per piece.
When story piece contains more story units, the \textbf{\textit{Portrait}} and \textbf{\textit{Landscape}} layouts are more suitable (\textcolor{black}{$M = 0.37$}), likely because these layouts offer larger space for each story piece (displayed across a full row or column). For a moderate density, the \textbf{\textit{Star}}, \textbf{\textit{Grid}}, and \textbf{\textit{PortraitGrid}} layouts are more appropriate ($M = 0.64$, $M = 0.72$, and $M = 0.70$, respectively), providing a balanced structure for organizing multiple story pieces. Conversely, when very few story units are embedded in each story piece, the \textbf{\textit{Spiral}} layout becomes more effective (\textcolor{black}{$M = 0.87$}). This might be due to the Spiral layout's property: story pieces are arranged along a continuous curve. Overloading each piece with too many story units would disrupt this reading flow. Additionally, we observed the overall distribution of story unit density across all infographics: \textcolor{black}{71.8\%} of story pieces contain only one story unit, \textcolor{black}{17.2\%} contain two, \textcolor{black}{5.6\%} contain three, and \textcolor{black}{3.4\%} contain four. Story pieces with five, six, or seven story units are extremely rare (\textcolor{black}{1.0\%, 0.8\%, and 0.2\%}, respectively). Given that story pieces containing more than four story units are uncommon, we limit each story piece to no more than four story units in our system to maintain readability.

\textbf{Proportion of Narrative Connections Between Story Pieces.}
We found that the proportion of narrative connections between story pieces plays an important role in shaping layout organization. 
A higher proportion indicates stronger interconnections between story pieces, while a lower proportion suggests that story pieces are more independent. Our analysis shows that when story pieces are loosely connected, the \textbf{\textit{Grid}} layout is more suitable (proportion = 28\%). For example, when the content consists of scattered story pieces, each piece individually related to the story goal but weakly connected to others.
When moderate logical connections exist, layouts such as \textbf{\textit{Star}}, \textbf{\textit{Portrait}}, \textbf{\textit{Landscape}}, and \textbf{\textit{PortraitGrid}} are more appropriate (proportions = 57\%, \textcolor{black}{52\%}, and 55\%, respectively), providing a balanced structure that supports both thematic grouping and independent pieces. For the \textbf{\textit{Spiral}} layout, the logical connections between story pieces tend to be slightly stronger (proportion = \textcolor{black}{72\%}), reflecting the continuous narrative flow guided by the spiral structure, where stronger interconnections between story pieces help maintain a smooth reading experience.

\textbf{Underlying Structuring Patterns.} We further observed that the narrative logic patterns between story pieces and the story goal, which reflect the underlying structuring strategies, are closely related to layout selection. Specifically, the \textbf{\textit{Portrait}} and \textbf{\textit{Landscape}} layouts often display a relatively balanced distribution of \textit{Generalization}, \textit{Elaboration}, and \textit{Example} relations. This pattern suggests that these layouts are well suited for a general-specific-detailed narrative structure, where the story begins with an overview, continues with elaborations on key aspects, and concludes with concrete examples. The \textbf{\textit{Star}} and \textbf{\textit{PortraitGrid}} layouts are more appropriate for a general-specific structure. In the \textbf{\textit{Star}} layout, a central visual graphic typically anchors the story, with surrounding story pieces maintaining strong semantic connections to the center, naturally forming a general-to-specific flow. Similarly, the \textbf{\textit{PortraitGrid}} layout exhibits a higher proportion of \textit{Generalization} and \textit{Example} relations, reinforcing this structure. Additionally, the \textbf{\textit{PortraitGrid}} layout shows a higher occurrence of \textit{Temporal} relations, supporting a sequential and time-based information flow.
The \textbf{\textit{Spiral layout}} is dominated by \textit{Example} relations between Story Pieces and the story goal, \textcolor{black}{with \textit{Temporal} relations forming the second most frequent type}, suggesting an instance-driven and \textcolor{black}{time-based} storytelling style.
The \textbf{\textit{Grid}} layout, by contrast, shows a greater diversity of narrative logic types across example cases.

\textbf{Gestalt Principle of Proximity.}
Based on the Gestalt Principles of Proximity \cite{koffka2013principles}, story pieces placed close to each other are perceived as more closely related than those positioned farther apart. Therefore, when designing the layout, it is important to position story pieces with narrative connections near one another to emphasize their logical relationships and enhance the reading experience. Additionally, we should ensure that visual designs belonging to the same story unit are grouped closely together, and that story units within the same story piece are also positioned near each other. 
\begin{figure*}[h]
    \centering
    \includegraphics[width=\linewidth]{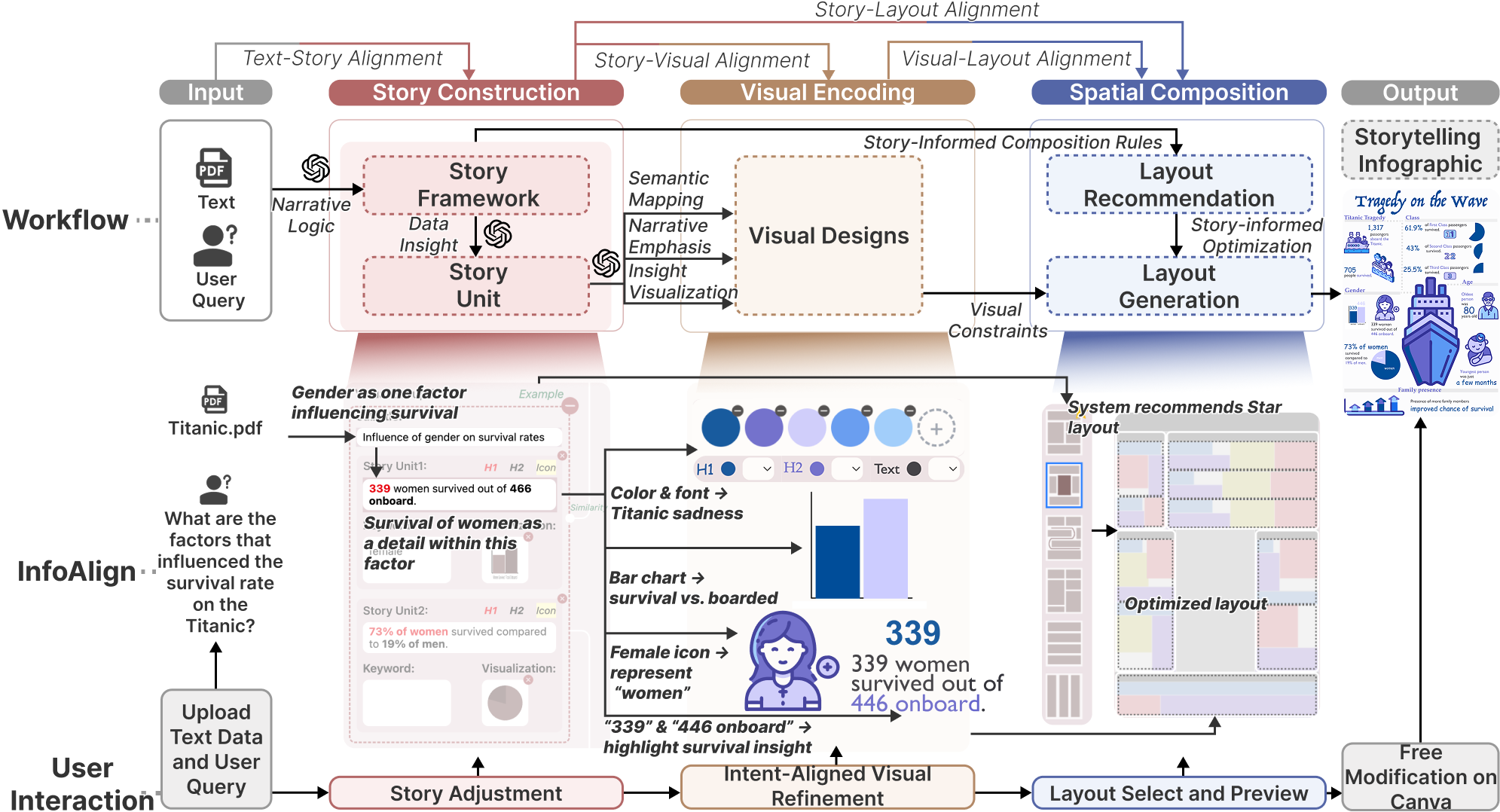}
    \caption{Overview of the narrative-centric workflow and its implementation in InfoAlign. The \textbf{Workflow} depicts the three phases of storytelling infographic creation: story construction, visual encoding, and spatial composition. The \textbf{InfoAlign} shows the system built on this workflow, demonstrated with a Titanic dataset example to illustrate how stories are consistently maintained across design steps.  The \textbf{User Interaction} highlights user interventions that incorporate story intent into each stage of the design process.}
    \label{fig:Workflow}
    \Description{Overview of the narrative-centric workflow and its implementation in InfoAlign. The Workflow depicts the three phases of storytelling infographic creation: story construction, visual encoding, and spatial composition. The InfoAlign shows the system built on this workflow, demonstrated with a Titanic dataset example to illustrate how stories are consistently maintained across design steps. The User Interaction highlights user interventions that incorporate story intent into each stage of the design process.}
\end{figure*}

\section{InfoAlign}
\label{sec-5}
Based on the summarized design requirements (\cref{sec-3.2}) and the derived story-connected layout patterns (\cref{sec-4.2}), we introduce a narrative-centric workflow for storytelling infographic generation. This workflow ensures that the story is consistently maintained across the design steps. Building on it, we propose InfoAlign, a human–AI co-creation system for generating storytelling infographics that preserves the user's story intent throughout the design process. 

\subsection{Narrative-Centric Workflow}
\label{sec-5.1}
The overview of the narrative-centric workflow is illustrated in \cref{fig:Workflow}-Workflow. The workflow consists of three phases: \textit{story construction}, \textit{visual encoding}, and \textit{spatial composition}. 
Given a long or unstructured text and a user query \textbf{(R1)}, the story is framed through narrative logic and extracting story units from data insights \textbf{(R2)}. Each story unit is then visually represented through semantic mapping, narrative emphasis, and insight visualization, producing candidate visual designs \textbf{(R3)}. Based on the constructed story and informed patterns, the system recommends layouts. Using both story patterns and visual constraints, a layout blueprint is generated to specify the positional arrangement of visual designs \textbf{(R4)}. Finally, by filling the visual designs into the recommended positions, the system produces the complete storytelling infographic. 

\textcolor{black}{Our workflow adopts a hybrid design that combines LLMs with rule-based algorithms. LLMs extract and structure story content from narrative-logic and data-insight instructions and generate icons from concise prompts. In contrast, layouts must satisfy coupled constraints based on the story patterns (e.g., reading order, SP/SU density, inter-piece connections), so we encode them as rule-based algorithms derived from corpus analysis to ensure coherent story flow, which current LLMs still struggle with due to limited spatial reasoning \cite{li2025llms} and mismatch of story structure \cite{lin2023layoutprompter}.}

\subsubsection{Story Construction}
\label{sec-5.2}
As shown in \cref{fig:Workflow}-Workflow, the process begins with story construction, where long or unstructured text is transformed into a structured story \textbf{(R1, R2)}. This phase ensures alignment between user input and the story.

\textbf{Structuring Story Framework From Input Using Narrative Logic.} To frame the story, we process the textual data into story pieces and connect them using narrative logic (defined in \cref{sec-4.1.2}). The LLM (GPT-4o mini) is instructed to act as a narrative extraction assistant that segments the input text into story pieces based on the user query and predefined narrative logic. We provide the model with detailed descriptions of each logic type, strict rules, and illustrative examples (Sec. A1 in the supplementary material). The model extracts story pieces relevant to the query, labels the narrative relations among them, and identifies potential relationships among pieces. It outputs the full content and title of each story piece, as well as their pairwise relations and their relations to the query. For example, in \cref{fig:Workflow}-InfoAlign, when uploading Titanic-related text with the query ``What factors influenced the survival rate on the Titanic?'', one extracted story piece is ``Influence of gender on survival rate'', which is an example story piece that answers the query.

\textbf{Extracting Story Units from Story Pieces Based on Data Insights.} To capture the detailed data-driven content within each story piece, we extract story units based on data insights (defined in \cref{sec-4.1.2}). We instruct the LLM (GPT-4o mini) to serve as an insight extraction assistant that identifies data-driven insights from each story piece. For this task, we provide detailed descriptions of each insight type, strict rules, and rich examples (Sec. A2 in the supplementary material). Given the full content of each story piece, the model extracts the most relevant story units that clearly and uniquely represent the underlying data insights. It outputs the content of each story unit along with its corresponding data insight. For example, in \cref{fig:Workflow}-InfoAlign, ``339 women survived out of 466 on board'' is one of the detailed data-driven story units within the story piece ``Influence of gender on survival rate''.

\subsubsection{Visual Encoding}
To align visual designs with the story, we represent each story unit through visual designs informed by semantic mapping, narrative emphasis, and insight visualization \textbf{(R3)}. The design forms include primary and secondary highlights, regular text, icons/graphics, and charts, as illustrated in \cref{fig:Workflow}-Workflow.

\textbf{Semantic Mapping.} The story can be transformed into visual representations through semantic mapping. Designs suitable for semantic mapping include \textbf{\textit{icons/graphics}} and \textbf{\textit{colors/fonts}}. We instruct the LLM (GPT-4o mini) to suggest a noun semantically related to each story unit (Sec. A2 in the supplementary material), which is then passed to RecraftAI to generate SVG graphics filled with the recommended colors. For colors and fonts, we further instruct the LLM to act as a professional palette and font generator, recommending color schemes and font styles based on the overall story semantics. We define the requirements for colors and fonts and provide explicit constraints to ensure semantic consistency (Sec. A5 in the supplementary material). For example, in \cref{fig:Workflow}-InfoAlign, a blue–purple color palette semantically reflects the sadness of the Titanic tragedy, while a female icon semantically reflects the story unit ``339 women survived out of 466 on board'' by highlighting the notion of ``women''.

\textbf{Narrative Emphasis.} To visually convey key story insights, we emphasize critical information through \textbf{\textit{primary highlights}} and \textbf{\textit{secondary highlights}}. We instruct the LLM (GPT-4o mini) to identify the most critical number or superlative as the primary highlight of each story unit, and to suggest contextual keywords as secondary highlights (Sec. A2 in the supplementary material). For example, in \cref{fig:Workflow}-InfoAlign, ``339'' in the story unit ``339 women survived out of 466 on board'' is the primary highlight, while ``466 on board'' serves as the secondary highlight, together emphasizing the insight into women's survival in the Titanic.

\textbf{Insight Visualization.} To present data-related information in each story unit, we offer a \textbf{\textit{chart}} representation of the contained insight when suitable, helping to intuitively reveal patterns, relations, or key figures. We instruct the LLM (GPT-4o mini) to assess whether a story unit is eligible for visualization based on its data insight type and content. Specifically, Pie Charts are used for Proportion or Difference with multiple entities; Bar Charts for Value, Difference, or Rank with multiple entities; Line Charts for Trend with sufficient time points; Single Pie Charts for Proportion with a single entity (e.g., showing a 61.9\% slice); and Pictographs for Proportion expressed as a fraction (e.g., ``1 in 10''). We provide the model with clear definitions of each data insight type, their corresponding chart mappings, strict rules, and illustrative examples (Sec. A2 in the supplementary material). The model output charts that suited the insight of the story unit. For example, in \cref{fig:Workflow}-InfoAlign, a bar chart shows the comparison of women who survived versus those on board the Titanic, corresponding to the story unit ``339 women survived out of 466 on board.'' \textcolor{black}{Charts are generated only when the story unit contains explicit numerical information (e.g., ``61.9\% of first class survived''). When the statement describes a qualitative or trend-based change without concrete values (e.g., ``improved chance of survival''), the system instead may suggest a metaphorical icon such as a upward arrow to express the direction of change. Although this abstraction may limit quantitative precision due to the lack of concrete numerical values in the input, it provides a visually accessible representation of qualitative trends.}

\subsubsection{Spatial Composition}
It is essential to retain a consistent story when arranging the infographic layout to guide the user's reading flow. As shown in \cref{fig:Workflow}-Workflow, we recommend infographic layouts through story-informed composition rules using a rule-based algorithm (based on \cref{sec-4.2}). To synchronize with the generated visual designs, we generate a layout blueprint that incorporates both story-driven and visual constraints through an optimized algorithm \textbf{(R4)}. This phase ensures alignment between story and layout, as well as between visual designs and layout.
 
\textbf{Layout Recommendation.} The layout is recommended based on rules informed by story patterns. Building on the findings from quantitative analysis (\cref{sec-4.2}), we developed a rule-based scoring algorithm that recommend candidate layouts. For each layout type $l$, a score $S(l)$ is incrementally adjusted according to several factors, including the number of story pieces, the ratio of story units to story pieces, the ratio of related story pieces, and the distribution of narrative logic. Additional adjustments are made based on cross-links among story pieces. Candidate layouts include \textit{Grid}, \textit{Spiral}, \textit{Landscape}, \textit{Star}, \textit{Portrait}, and \textit{PortraitGrid} (see \cref{fig:Analysis}). After computing the scores, the system ranks all layouts from the most to the least recommended. Let the set of candidate layouts be
\begin{equation}
\label{eq:layouts}
    \mathcal{L} = \{ \text{Grid}, \text{Spiral}, \text{Landscape}, \text{Star}, \text{Portrait}, \text{PortraitGrid} \}.
\end{equation}

For each layout $l \in \mathcal{L}$, we define its score as
\begin{equation}
\label{eq:score}
    S(l) = \sum_{f \in \mathcal{F}} \delta_f(l) + \delta_{\text{ref}}(l),
\end{equation}
where $\mathcal{F}$ denotes the set of story-informed factors (e.g., number of story pieces $N_{SP}$, 
ratio of story units $\rho_{SU} = N_{SU}/N_{SP}$, 
ratio of related story pieces $\rho_{rel} = N_{rel}/N_{SP}$, 
and narrative logic distribution $R = \{r_i\}$). 

Each $\delta_f(l)$ is a rule-based increment:
\begin{equation}
\label{eq:delta}
\delta_f(l) =
\begin{cases}
1, & \text{if the condition for layout $l$ is satisfied}, \\
0, & \text{otherwise}.
\end{cases}
\end{equation}

For example:
\begin{equation}
\label{eq:examples}
\begin{aligned}
&N_{SP} > 8 \;\Rightarrow\; S(\text{Grid}){+}{=}1, \\
&0.3 \leq \rho_{SU} \leq 0.6 \;\Rightarrow\; S(\text{Star}){+}{=}1, \\
&\rho_{rel} > 0.8 \;\Rightarrow\; S(\text{Spiral}){+}{=}1, \\
&\#\{r_i > 0\} > 3 \;\Rightarrow\; S(\text{Portrait}){+}{=}1, \\
&\max(\text{ref}) > 2 \;\Rightarrow\; S(\text{PortraitGrid}){+}{=}1.
\end{aligned}
\end{equation}

Here, $N_{SP}$ denotes the number of story pieces, 
$N_{SU}$ the number of story units, 
$N_{rel}$ the number of related story pieces, 
$\rho_{SU}$ and $\rho_{rel}$ their corresponding ratios, 
$r_i$ the count of the $i$-th narrative relation type in $R$, 
and $\text{ref}$ the in-degree count vector of story pieces 
(i.e., how many times each story piece is linked from others).

The layouts are then ranked as
\begin{equation}
\label{eq:ranking}
    [l_1, l_2, \dots, l_6] = \text{sort}\{S(l) \mid l \in \mathcal{L}\}.
\end{equation}

\textbf{Layout Generation.} We then propose an optimized algorithm to generate layout blueprint that informed by both story and visual constraints based on the recommended layout.
Generally, the algorithms for different layout formats adhere to a common framework, with additional steps incorporated for the Spiral and Star formats. Within this framework, the algorithm starts by determining the placement of SPs based on the number of original SPs\( \mathcal{V} \), customized for each layout format as specified in Figure~\ref{fig:Analysis}. In the infographic, the title is placed at the top, spanning the full width \( W \), while each SP consists of a subtitle followed by its SUs stacked vertically.

Prior to calculating precise positions and sizes, we establishes soft constraints for the algorithm:
\begin{itemize}
    \item Let the height of the letters in the main text be \( x \). Then the letter height in the title is \( 3x \), with SP subtitles at \( 1.5x \), and highlight text at \( 2x \).
    \item The ideal area ratio for the icon, chart, and main text within a SP is 1:1:1.
\end{itemize}
These constraints serve as optimization targets, though adjustments may be made to accommodate element overlaps.

To compute the widths, heights, and positions of SPs, SUs, and visual designs, the total infographic area is expressed as a function of \( x \) and equated to the fixed area:
\begin{equation}
\label{eq:main}
    A(x) = W H,
\end{equation}
where H is the height of the infographic, and \( A(x) = A_{\text{title}}(x) + \sum_{\text{SP} \in \mathcal{V}} A_{\text{SP}}(x) \). Each SP's area is:
\begin{equation}
    A_{\text{SP}}(x) = A_{\text{subtitle}}(x) + \sum_{\text{SP} \in \text{SP}} A_{\text{SU}}(x),
\end{equation}
with each SU's area defined as:
\begin{equation}
    A_{\text{SU}}(x) = A_{\text{highlight}}(x) + A_{\text{text}}(x) + \mathbf{1}_{\text{icon}} A_{\text{icon}}(x) + \mathbf{1}_{\text{chart}} A_{\text{chart}}(x).
\end{equation}

The individual area components are:
\begin{align}
    A_{\text{title}}(x) &= 3 W x, \\
    A_{\text{subtitle}}(x) &= 1.5 W_{\text{SP}} x, \\
    A_{\text{highlight}}(x) &= 2 x \cdot 2W_{\text{text}}(x, \text{highlight}), \\
    A_{\text{text}}(x) &= x \cdot W_{\text{text}}(x, \text{text}), \\
    A_{\text{icon}}(x) &= A_{\text{chart}}(x) = A_{\text{text}}(x),
\end{align}
where \( W_{\text{SP}} \) is the width allocated to the SP, and \( W_{\text{text}}\) is the spatial length of a piece of text at height \( x \). The text length \( W_{\text{text}}\) is derived from the text height and the width-to-height ratios.

Solving Equation~\ref{eq:main} yields \( x \) in pixels, enabling the computation of SP, SU, and visual designs. SP positions, widths, and heights are then assigned based on their areas and the layout format's constraints, with SU dimensions similarly derived within each SP.

For icons and charts, we employ a set of height-to-width ratios and define corresponding placement strategies. The algorithm iterates through combinations of ratios and strategies for each SU, seeking positions, widths, and heights that avoid overlaps with the subtitle, icon, and chart, while staying within SU boundaries. The main text occupies the largest remaining rectangular block, determined by the selected placement. If no combination eliminates overlaps, the algorithm selects the option with minimal overlap and resizes the icon and chart by the smallest margin to resolve it.

The Spiral and Star layouts extend the framework with:
\begin{itemize}
    \item \textbf{Spiral}: SPs are reordered to align with the narrative logic in Figure~\ref{fig:Analysis}. SPs enclosed by the curve from three sides are resized smaller.
    \item \textbf{Star}: A virtual SP with an area of \( \frac{1}{4} \) the total SP content area is added to \( \mathcal{V} \).
\end{itemize}

Ultimately, our algorithms provide the layout blueprint for each story pieces, story units, and visual designs.

\begin{figure*}[t]
    \centering
    \includegraphics[width=\linewidth]{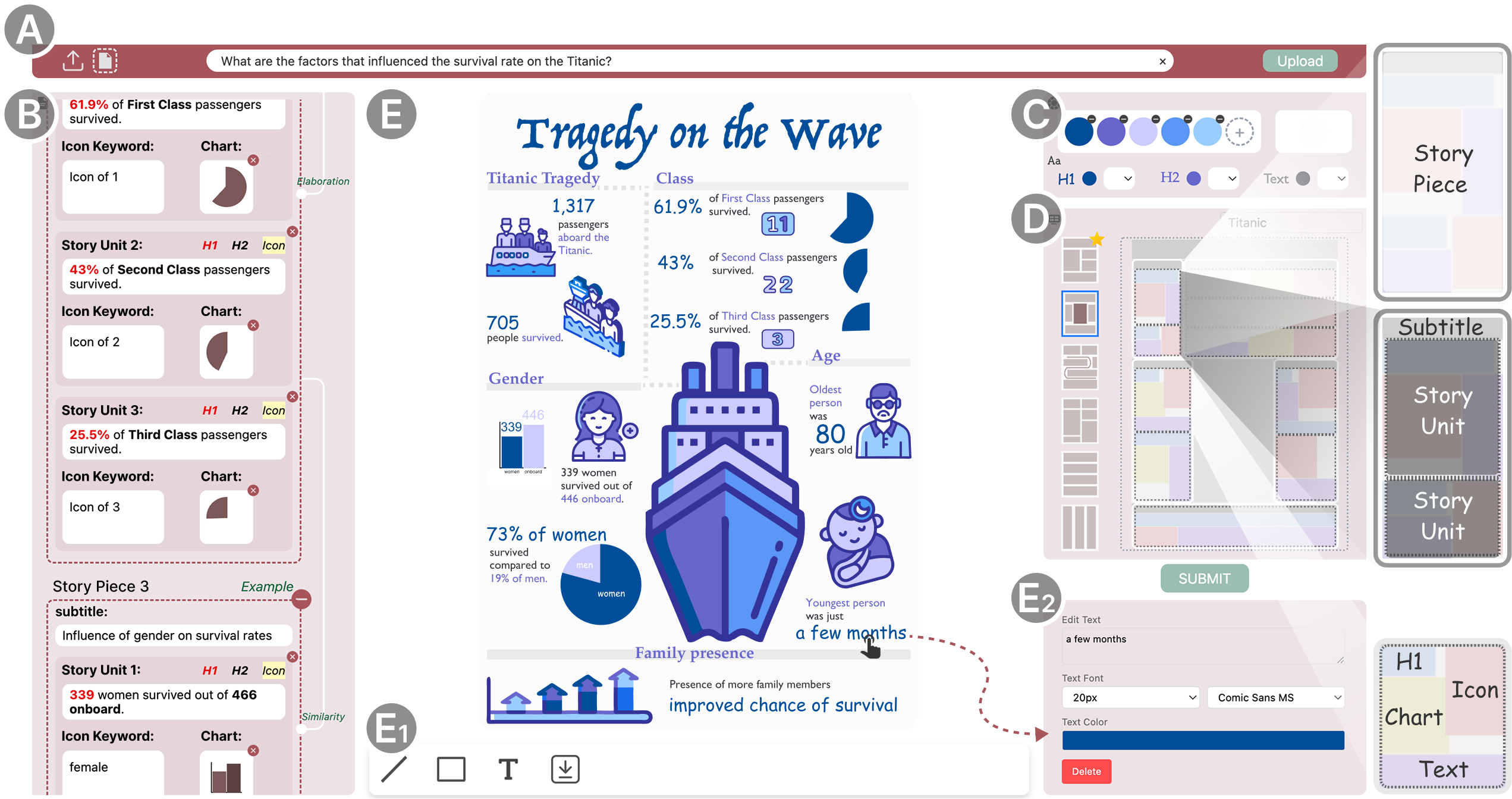}
    \caption{%
    The InfoAlign system consists of five interactive views: 
    (A) the Input View, which enables document upload and query definition; 
    (B) the Story View, which constructs stories and recommends visual designs; 
    (C) the Stylization View, which suggests colors and fonts; 
    (D) the Layout View, which offers ranked layout blueprints; and (E) the Canva View, which displays the final storytelling infographic with interactive customization options (including freeform annotation (E1) and text editing (E2)).}
    \label{fig:System}
    \Description{The InfoAlign system consists of five interactive views: (A) the Input View, which enables document upload and query definition; (B) the Story View, which constructs stories and recommends visual designs; (C) the Stylization View, which suggests colors and fonts; (D) the Layout View, which offers ranked layout blueprints; and (E) the Canva View, which displays the final storytelling infographic with interactive customization options (including freeform annotation (E1) and text editing (E2)).}
\end{figure*}

\subsection{System}
\label{}
Based on the narrative-centric workflow, we developed InfoAlign, a human–AI collaboration system that supports storytelling infographic generation, allowing users to manually intervene and adjust the design at each step \textbf{(R5)}, thereby ensuring their story intent is maintained throughout the creation process. InfoAlign provides five views, designed to guide users through the process of transforming textual data and user queries into storytelling infographics.

\textbf{Input View.}
The Input View (\cref{fig:System}A) enables users to upload unstructured and lengthy textual data and define their story goal (\textbf{R1}). Users first upload a document (e.g., a PDF) and specify a user query (e.g., ``What are the factors that influenced the survival rate on Titanic?''). Clicking ``Upload'' submits the input and initiates the structuring process.

\textbf{Story View.}
The Story View (\cref{fig:System}B) assists users in automatically constructing stories by organizing story pieces (SPs) and their corresponding story units (SUs) \textbf{(R2)}. Each SP is assigned a subtitle (e.g., ``Influence of gender on survival rate'') and annotated with its narrative relationship to the user query (e.g., ``example''). Logical relations between SPs (e.g., a ``similarity'' relation between ``class'' and ``gender'') are visually indicated. Within each SP, SUs are contained, each associated with content and visual designs \textbf{(R3)}. For every SU, the system automatically generates primary highlights (bold red) and secondary highlights (bold black), along with suggested keywords for icon/graphic generation and appropriate charts. Users retain full flexibility to manually adjust the constructed stories \textbf{(R5)}. They can refine or delete SPs (e.g., remove an unwanted piece, modified story content), adjust highlights, modify or remove charts, and edit icon keywords (e.g., change the keyword ``female'' to ``women''). This view ensuring that both the story and its visual designs align with user intent.

\textbf{Stylization View.}
The Stylization View (\cref{fig:System}C) recommends color schemes and fonts aligned with the semantic and emotional tone of the story \textbf{(R3)}. The system suggests 3–5 theme colors and one background color, along with fonts and colors for primary highlights, secondary highlights, and regular text. All stylistic options are editable, allowing users to customize them to meet personal requirements (e.g., changing blue to green) \textbf{(R5)}.

\textbf{Layout View.}
The Layout View (\cref{fig:System}D) recommends candidate infographic layouts based on the story pattern \textbf{(R4)}. It ranks six templates (Grid, Star, Portrait, Landscape, PortraitGrid, and Spiral) according to their suitability. Users can preview how the story is mapped into each layout, which allows them to assess the overall flow and compare alternatives before making a decision. Layout selection is interactive \textbf{(R5)}, users can browse different options, modify the story if needed, and re-select the layout by confirming their choice (e.g., selecting Star instead of Grid).

\textbf{Canva View.}
The Canva View (\cref{fig:System}E) provides an interactive canvas where the generated infographic is rendered as editable SVG elements. Users can manipulate visual designs directly by dragging elements to reposition them and adjusting their spatial arrangements. Selecting a visual design (e.g., highlighted text such as ``a few months'') opens detailed editing options (\cref{fig:System}E2), including content modification, font adjustments, and color changes. In addition, freeform canva enable users to draw lines and rectangles (\cref{fig:System}E1) or add new text annotations, offering flexibility for further personalization \textbf{(R5)}. After refinement, the final infographic can be exported as a high-quality SVG file using the save function.

\section{User Study}
\label{sec-6}
Because InfoAlign is a human–AI co-creation system built around a narrative-centric workflow for storytelling infographic creation, our evaluation examines both the quality of the workflow and how users engage with the system across its design stages. We therefore focused on three dimensions: (1) workflow-level quality, assessing whether story construction, visual encoding, and spatial composition remained coherent and aligned with users' story goals; (2) human–AI interaction patterns, characterizing how participants modified, adopted, or refined system outputs during creation; and (3) overall system effectiveness, capturing usability, creativity support, perceived co-creation, and the visual aesthetics of the resulting infographics. We conducted the study with 12 infographic creators, the results are reported in \cref{sec-7}.

\subsection{Participants}
We recruited 12 participants (U1–U12) via snowball sampling. Inclusion criteria required participants to be over 18, with a good command in English, and experienced in infographic design. The sample included six males and six females, aged 20 to 29 years ($M = 23.42$, $SD = 2.64$). None had participated in earlier formative study. All participants reported familiarity with storytelling infographics and prior exposure to such designs. Regarding design experience, one had five years, one had four years, three had three years, five had two years, and two had one year of relevant infographic design experience. Two participants majored in design, four in media and communication, and the remaining in computer science. All had experience using GPT-based tools for summarizing or analyzing data. Participants received \$8 as compensation.

\subsection{Study Design}
We adopted a mixed-method, task-based study in which participants used InfoAlign to complete an end-to-end storytelling infographic while thinking aloud. This setup allowed us to examine InfoAlign's narrative-centric workflow and its support for human–AI co-creation across design stages by observing how system-generated outputs preserved story consistency, how participants intervened to refine them, and how effective the overall authoring experience was. During the study, we logged interactions and time spent in each view, collected stage-wise Likert ratings after automatically generated results, and administered a post-task questionnaire and semi-structured interview.

\begin{figure*}[p]
    \centering
    \includegraphics[width=0.81\linewidth]{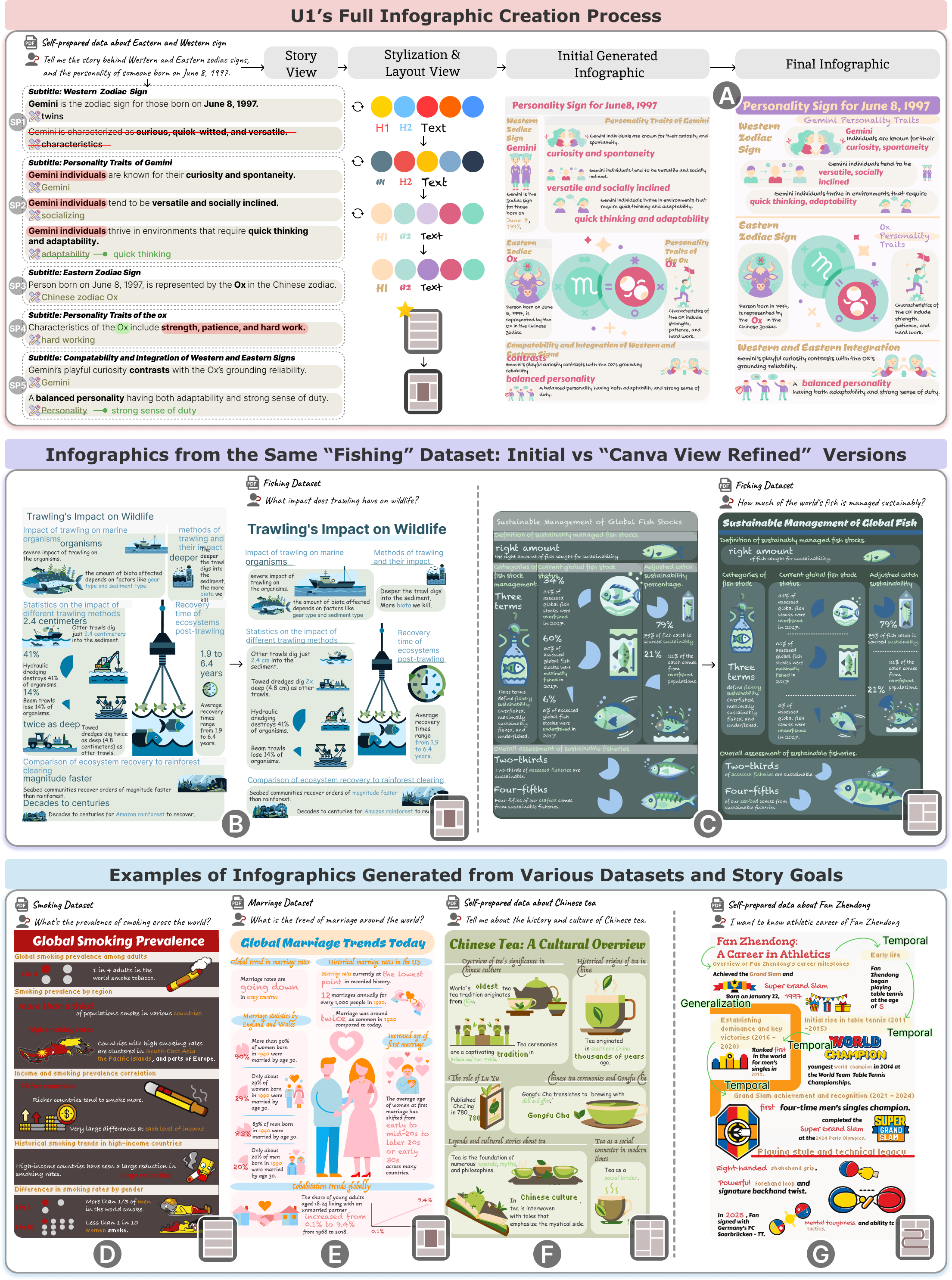}
    \caption{\protect\textcolor{black}{Infographics created with InfoAlign in the user study. (A) U1's full design process for a self-curated ``Western and Eastern zodiac'' dataset, showing how the system-generated storyframe, stylization, and layout are refined into a final infographic across the four views. (B–C) Results from the same ``Fishing'' dataset, each contrasting the initial system output with the Canva-refined version for different story goals. (D–G) Examples from the ``Smoking'' dataset (D), the ``Marriage and Divorce'' dataset (E), a self-curated ``Chinese tea'' dataset (F), and a self-curated ``Fan Zhendong'' dataset (G), with narrative logic annotated in (G).}}
    \label{fig:UserStudyGallery}
    \Description{Infographics created with InfoAlign in the user study. (A) U1’s full design process for a self-curated “Western and Eastern zodiac” dataset, showing how the system-generated storyframe, stylization, and layout are refined into a final infographic across the four views. (B–C) Results from the same “Fishing” dataset, each contrasting the initial system output with the Canva-refined version for different story goals. (D–G) Examples from the “Smoking” dataset (D), the “Marriage and Divorce” dataset (E), a self-curated “Chinese tea” dataset (F), and a self-curated “Fan Zhendong” dataset (G), with narrative logic annotated in (G).}
\end{figure*}

\subsubsection{Tasks and Materials} 
\textcolor{black}{Participants were asked to create a storytelling infographic using InfoAlign based on a textual dataset and a self-defined story goal. Four prepared datasets were provided, spanning the following topics: \textit{Marriage and Divorce} (2116 words), \textit{Smoking} (2152 words), \textit{Fishing} (1776 words), and \textit{Titanic} (10,326 words). Participants could also use a self-curated dataset, and four participants (U1, U3, U4 and U9) chose this option. Because most participants were unfamiliar with the prepared datasets, they first used \textcolor{black}{GPT-5} to explore and understand the content before starting the creation task. This familiarization step was skipped by those using self-curated datasets.}

\subsubsection{Procedure}
\textcolor{black}{The study consisted of four phases: introduction (10 min), dataset familiarization (10 min), infographic creation (no time limit), and questionnaire and interview (20 min). All sessions were conducted in a quiet environment. Participants used a desktop computer to interact with InfoAlign and a secondary monitor to complete stage-wise ratings and questionnaires.}

\textcolor{black}{In the introduction, participants viewed a short demo video and received a brief verbal explanation of InfoAlign. During dataset familiarization, they explored the assigned dataset using \textcolor{black}{GPT-5} unless working with self-curated material. In the infographic creation phase, participants completed InfoAlign's full workflow. At each stage, the system first generated an initial result, which participants rated using a short 7-point Likert questionnaire before freely refining the output. They progressed through the \textit{Input View}, reviewing and adjusting the storyframe and visual designs in the \textit{Story View}, examining and optionally modifying system-recommended styles in the \textit{Stylization View}, evaluating and browsing layouts in the \textit{Layout View}, and finalizing the infographic in the \textit{Canva View}. Participants verbalized their thoughts throughout. With consent, both screen activity and audio were recorded. After completing the infographic, participants filled out a post-task questionnaire and participated in a semi-structured interview.}

\newcolumntype{L}[1]{>{\RaggedRight\arraybackslash}m{#1}}
\renewcommand{\arraystretch}{1.3}
\setlength{\tabcolsep}{8pt}

\begin{table*}[h!]
\centering
\caption{\protect\textcolor{black}{Workflow-level Evaluation questionnaire items grouped by three dimensions. All items were rated on a 7-point Likert scale (1 = strongly disagree, 7 = strongly agree).}}
\label{tab:stepquestionnaire}
\begin{tabular}{@{}p{0.22\linewidth} p{0.74\linewidth}@{}}
\toprule
\textbf{Dimension} & \textbf{Questionnaire Items} \\
\midrule
\multirow{4}{*}{\makecell[l]{\textbf{Story}\\\textbf{Construction}}}
 & Q1. The role of this story piece in the overall story is clear and makes sense. \\
 & Q2. The connections between this story piece and its related story pieces are clear.\\
 & Q3. This story unit clearly elaborates the idea of its corresponding story piece. \\
 & Q4. The overall storyframe is in line with my story goal. \\
\midrule
\multirow{4}{*}{\makecell[l]{\textbf{Visual}\\\textbf{Encoding}}}
 & Q5. The icon keyword for this story unit semantically matches its content. \\
 & Q6. The chart for this story unit appropriately visualizes its data or insight. \\
 & Q7. The highlighted text in this story unit correctly emphasizes the key information. \\
 & Q8. The colors and fonts are consistent with the theme and tone of the story. \\
\midrule
\multirow{3}{*}{\makecell[l]{\textbf{Spatial}\\\textbf{Composition}}}
 & Q9. The layout presents the story in a clear and logical sequence. \\
 & Q10. The overall narrative flows smoothly from beginning to end. \\
 & Q11. The layout is well-structured. \\
\bottomrule
\end{tabular}
\Description{Workflow-level Evaluation questionnaire items grouped by three dimensions. All items were rated on a 7-point Likert scale (1 = strongly disagree, 7 = strongly agree).}
\end{table*}

\subsubsection{Measures}
We collected both quantitative and qualitative measures aligned with our three evaluation goals.

\textbf{Workflow-Level Quality.}
To assess the quality of InfoAlign's narrative-centric workflow, we captured participants' stage-wise ratings of how coherent\cite{graesser2004coh,  green2000role} and consistent\cite{reese2011coherence} the system-generated story frame, visual designs, and layouts were with the overall story(\cref{tab:stepquestionnaire}). After automatically generated result in the \textit{Story View}, participants were asked to answer Q1–Q7 on 7-point Likert scales. For Q1, Q2, Q3, Q5, Q6, and Q7, they rated each generated component (e.g., each story unit) individually. Participants then completed Q8 (7-point Likert) after the generation in the \textit{Stylization View}, and Q9–Q11 (7-point Likert) after the generated result in the \textit{Layout View}. To quantify the factual reliability of the generated story units, two coders, both fluent in English and familiar with the task and not involved in the user study, independently evaluated each story unit against the original dataset. Each unit was labeled as Correct (faithfully expressing facts present in the data), Incorrect (contradicting or distorting the data), or LLM-Extended (not explicitly stated yet inferable from the context). Inter-rater agreement was substantial, with a Cohen's $\kappa$ of 0.742.

\textbf{Human–AI Interaction Patterns.}
To capture interaction patterns and the degree of user intervention during infographic creation, we derived logged interaction data from screen recordings. Specifically, we counted the frequency of modifications and deletions in the \textit{Story View}, the number of refresh actions and whether participants adjusted the stylizations in the \textit{Stylization View}, and the number of layout-browsing actions and whether participants selected the system-recommended layout in the \textit{Layout View}. In the \textit{Canva View}, interactions spanned many fine-grained operations (e.g., resizing, rotating, adding or removing elements), making it difficult to code them reliably. As a complementary measure, we recorded the time participants spent in each view, including the \textit{Canva View}. The reported time duration reflects only the time participants actively interacted with the system, excluding any time spent completing questionnaires.

\textbf{System Effectiveness.}
After completing the final infographic, participants completed standardized 7-point Likert questionnaires (from ``strongly disagree'' to ``strongly agree'') assessing four aspects of the system. Usability was measured using a subset of the SUS~\cite{brooke1996sus}, creativity support using the CSI~\cite{cherry2014quantifying}, perceived human-AI co-creation using items adapted from MICSI~\cite{lawton2023drawing}, and perceived visual aesthetics using items adapted from VisAWI~\cite{moshagen2010facets}. (See Sec. D in the supplementary material for the full item wording.)

\subsubsection{Study Limitation}
A methodological consideration is that participants provided stage-wise ratings during the creation process, which might have introduced slight interruptions to their natural creation flow. These ratings were directly tied to the outputs shown at each step and integrated into the workflow, and a secondary monitor helped reduce attention shifts when switching interfaces. The evaluations were reasonably aligned with the ongoing tasks, this potential influence on how participants engaged with the creation process should be kept in mind when interpreting the results.

\begin{figure*}[h]
    \centering
    \includegraphics[width=\linewidth]{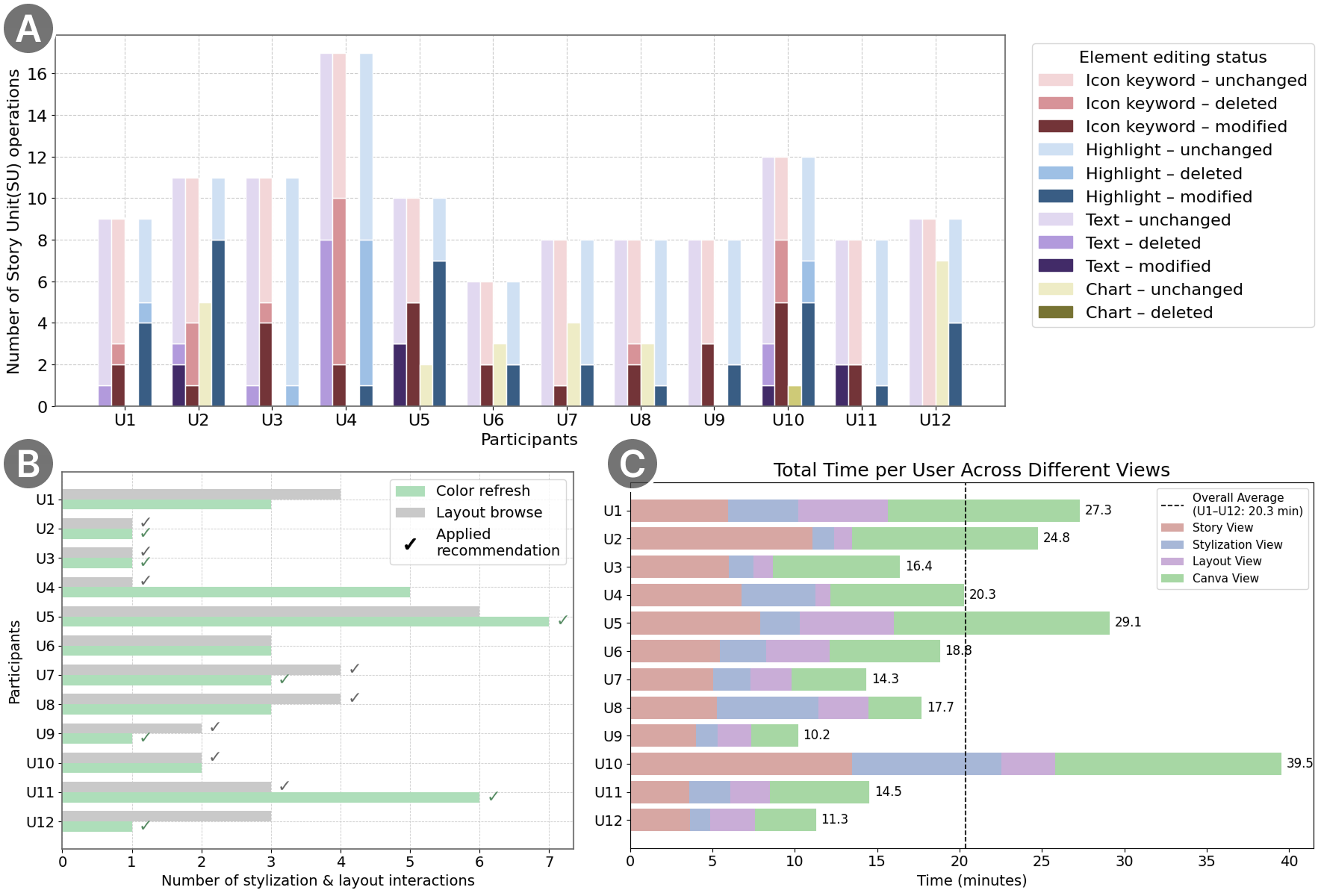}
    \caption{\protect\textcolor{black}{Interaction patterns during human–AI co-creation with InfoAlign. (A) Story-unit (SU) operations per participant (U1–U12) in the Story View, counting for each icon keyword, highlight, text block, and chart whether it was kept, modified, or deleted. (B) Stylization and layout interactions per participant, including the number of color-refresh, layout-browse actions, and whether system-recommended styles or layouts were applied. (C) Stacked bar chart of total creation time per participant across the four views (Story, Stylization, Layout, and Canva); the dashed line marks the overall average of 20.3 minutes.}}
    \label{fig:InteractionResult}
    \Description{Interaction patterns during human–AI co-creation with InfoAlign. (A) Story-unit (SU) operations per participant (U1–U12) in the Story View, counting for each icon keyword, highlight, text block, and chart whether it was kept, modified, or deleted. (B) Stylization and layout interactions per participant, including the number of color-refresh, layout-browse actions, and whether system-recommended styles or layouts were applied. (C) Stacked bar chart of total creation time per participant across the four views (Story, Stylization, Layout, and Canva); the dashed line marks the overall average of 20.3 minutes.}
\end{figure*}

\begin{figure*}[h]
    \centering
    \includegraphics[width=\linewidth]{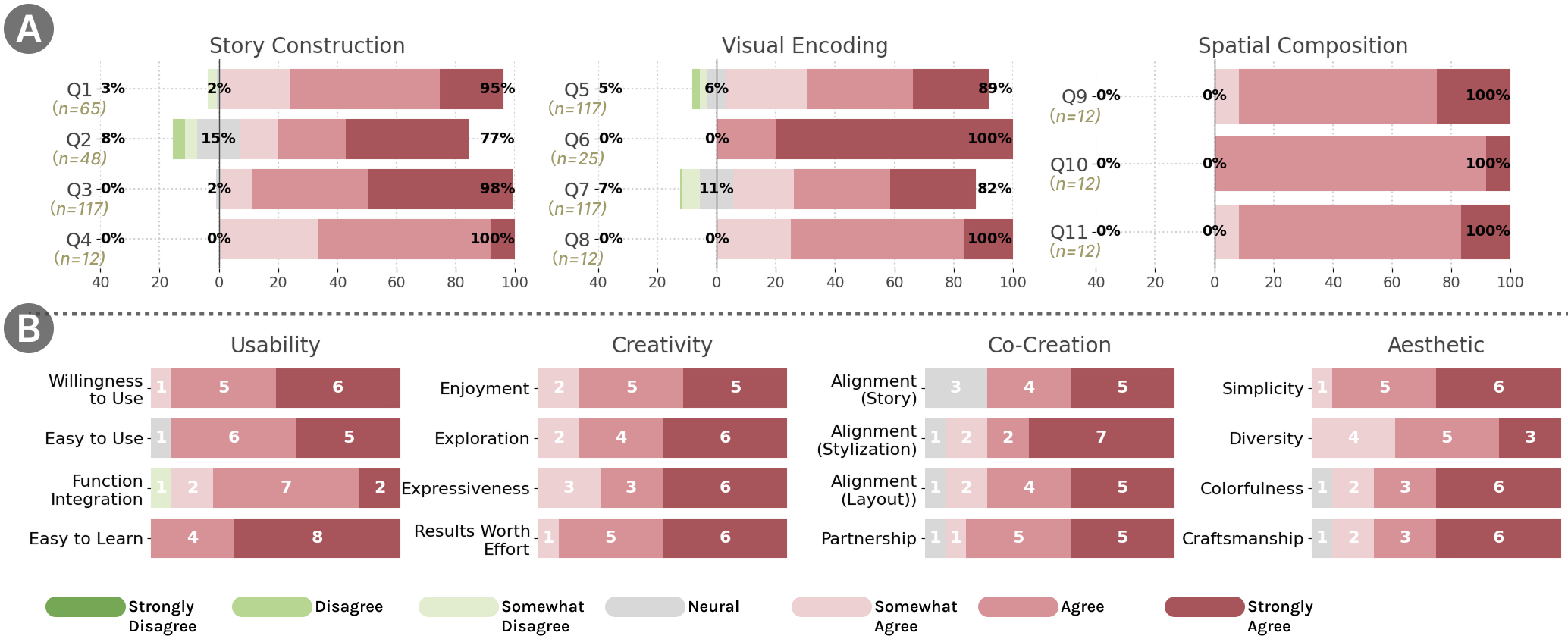}
    \caption{\protect\textcolor{black}{Quantitative evaluation of InfoAlign. (A) Workflow-level quality ratings for story construction, visual encoding, and spatial composition. Each stacked bar summarizes responses on a 7-point Likert scale (1 = strongly disagree, 7 = strongly agree); yellow labels show the number of samples, and the vertical line marks the neutral midpoint, with percentages to the left and right indicating responses below and above 4. (B) Overall effectiveness ratings for usability, creativity, co-creation, and aesthetics on the same scale.}}
    \label{fig:SystemResult}
    \Description{Quantitative evaluation of InfoAlign. (A) Workflow-level quality ratings for story construction, visual encoding, and spatial composition. Each stacked bar summarizes responses on a 7-point Likert scale (1 = strongly disagree, 7 = strongly agree); yellow labels show the number of samples, and the vertical line marks the neutral midpoint, with percentages to the left and right indicating responses below and above 4. (B) Overall effectiveness ratings for usability, creativity, co-creation, and aesthetics on the same scale.}
\end{figure*}
\section{Results}
\label{sec-7}
The results of the user study are analyzed as follows.

\subsection{Workflow-Level Story Consistency Across Creation Stages}
\label{sec-7.1}
We assessed workflow-level quality by examining whether InfoAlign maintained a consistent story across story construction (Q1-Q4), visual encoding (Q5-Q8), and spatial composition (Q9-Q10), as captured in \cref{tab:stepquestionnaire}.

\subsubsection{Aligning Story Goals and Constructed Stories}
Overall, participants felt that the system-constructed stories were clear in structure and well aligned with their story goals.
Across the 65 generated story pieces, 95\% were rated as coherent with the story goal (Q1), indicating a clear storyline.
As U3 noted, ``the pieces aligned with my story goal, and the logical connections made them easy to follow.'' 
Among the 48 logical relations between story pieces, 77\% were rated as clear (Q2), which is slightly lower than the ratings for other aspects. This lower score may reflect subjective interpretations of how story pieces should be related. For example, U1 (\cref{fig:UserStudyGallery}A) felt that the Western and Eastern zodiac signs are parallel and therefore should be treated as ``similar'' rather than ``contrast.'' Nonetheless, several participants appreciated having these connections made explicit; as P7 noted, ``these relationships help me see why each piece is there and how the story develops, and they increase my trust in the storyframe.''

\textcolor{black}{The elaborated story units of each story piece was also positively received. Across 117 story units, 98\% were rated as coherent with their corresponding story piece (Q3). This indicate good quality of filling the story in concrete details. A factual accuracy check further confirmed the reliability of the generated content: 95.73\% of story units accurately reflected the source data, 0\% contained incorrect statements, and 4.27\% were categorized as reasonable LLM-Extended inferences. For example, in U1's case (\cref{fig:UserStudyGallery}A), the sentence `a balanced personality having both adaptability and a strong sense of duty' is not stated directly in the dataset. However, both coders judged it to be a reasonable inference that synthesizes the description of Gemini's `adaptability' and the Ox's `strong sense of duty' into a `balanced personality' summary, consistent with U1's story goal of understanding personality; U1 also expressed satisfaction with this conclusion. Taken together, all participants agreed that the overall storyframe was consistent with their provided story goals (Q4). This suggests that InfoAlign first establishes a story backbone, then clarifies how the story develops from one part to the next, and finally fills in concrete details, resulting in a coherent storyframe that remains aligned with the story goal.}

\subsubsection{Story-Aligned Visual Design}
The system's visual encodings, including icons, charts, highlighted text, and color and font stylization, were generally perceived as consistent and coherent with the constructed stories. Among 117 suggested icons, 89\% were rated as semantically appropriate for their corresponding story details (Q5). Participants also noted that the automatic suggestions were helpful and sometimes even inspiring. U5 (\cref{fig:UserStudyGallery}B), for example, appreciated the recommendation of a ``clock'' icon for describing ``average recovery times,'' commenting, ``It was a pleasant surprise.'' A few participants also mentioned occasional visual homogeneity. U4 (\cref{fig:UserStudyGallery}G) noted that medal-related icons appeared too frequently, making the visuals feel repetitive despite being semantically appropriate for a story about frequent awards. All 25 recommended charts were rated as appropriately reflecting the underlying data insights (Q6). Participants found both the chosen chart types and the information encoded in them suitable for visually representing the data. Highlighting quality was similarly high. Among 117 recommended highlights, 98\% were judged as correctly emphasizing key information in the story units (Q7), indicating that the system reliably captured salient information within each story detail. All participants agreed that the system's initial color and font stylizations were consistent with the thematic tone of their stories (Q8). U12 (\cref{fig:UserStudyGallery}C) remarked, ``the recommended colors are appealing and clearly reflect the fishing theme.'' Overall, participants praised the quality of the automatically generated visual designs and emphasized the importance of visual presentation. U9 (\cref{fig:UserStudyGallery}F) noted, ``the visuals are what first attract attention and make the story understandable in an infographic, and InfoAlign's suggestions give me a good starting point.''

\subsubsection{Story-Informed Spatial Composition for Coherent Reading Flow}
All system-recommended layouts were perceived as well structured (Q11), presenting the story in a logical sequence (Q9) with a smooth reading flow (Q10). These results indicate that the layouts effectively integrated the structured storyframe and visual elements into a unified presentation whose reading order aligned with the story. For example, U4 (\cref{fig:UserStudyGallery}G) remarked, ``The spiral layout first summarizes Fan Zhendong (a famous table tennis player)'s career at the top-left, then unfolds his milestones along the spiral curve in temporal order, and finally closes with examples of his playing habits; this layout makes the story feel smooth, logical, and easy to follow.'' Several participants noted that generating such a layout manually would have taken significant time and effort. Overall, participants emphasized that InfoAlign's spatial composition reliably maintained narrative coherence in the final infographic. U2 (\cref{fig:UserStudyGallery}E) remarked that ``layout is central to ensuring the final storytelling infographic remains complete, and InfoAlign's recommended layout is clean and matches the story, nothing feels disconnected.''

\subsection{Human–AI Interaction Patterns Across Authoring Stages}
\label{sec-7.2}
To examine human–AI co-creation dynamics, we analyzed participants' interaction logs across the \textit{Story}, \textit{Stylization}, and \textit{Layout} views (\cref{fig:InteractionResult}A,B), along with time spent in each view (\cref{fig:InteractionResult}C).

\subsubsection{Integrating Story Goals and Creator Intent through Human–AI Co-Creation}
\label{sec-7.2.1}
\textcolor{black}{We analyzed interaction logs in the \textit{Story View}(\cref{fig:InteractionResult}A) and found that no participants removed any story pieces, so we focused on interactions within each story unit. We computed an adjustment rate for each element type in the story units, defined as the proportion of system-generated elements that were either modified or deleted by participants. Overall, adjustment rates were highest for highlights (41.0\%) and icons (39.3\%), lower for story text (17.9\%), and minimal for charts (4.0\%). Given that the workflow-level evaluation already showed high initial appropriateness for icons (89\% semantic match) and highlights (98\% emphasis correctness), participants frequently modified these elements not because they were incoherent with the story, but because they did not fully reflect their narrative goal. For instance, in U1's case (\cref{fig:UserStudyGallery}A), U1 agreed that ``strength, patience, and hard work'' correctly emphasized the key messages, yet removed these highlights to instead foreground ``Ox,'' explaining, ``I personally would want to emphasize Ox.'' Some participants selectively adjusting the story scope. U4 deleted five entire story units, along with their associated text, icons, and highlights, explaining that ``too many early career details are described, it's not the primary story I want to present.'' In general, participants made few changes to textual content, except for U5 (\cref{fig:UserStudyGallery}B), who was strict about wording and format. Modifications to charts were even rarer: only U10 deleted one chart. These patterns suggest that participants were more inclined to intervene in expressive visual designs, such as icons and highlights, to convey personal intent and subjective style, while largely accepting structured and statistical recommendations, such as the system-generated story text and charts. Because these elements were perceived as relatively objective and tightly linked to factual accuracy and subsequent story communication, most participants only made local adjustments within the recommended structure, rather than reshaping them as freely as they did with icons and highlights.}

\textcolor{black}{We further analyzed interactions in the \textit{Stylization View} (\cref{fig:InteractionResult}B) and \textit{Layout View} (\cref{fig:InteractionResult}C). In the \textit{Stylization View}, eight participants refreshed style suggestions multiple times and five refined the recommended styles. Given that all participants had already rated the initial styles as consistent with the story theme, these edits mainly reflected aesthetic preferences. An interesting pattern emerged. Instead of choosing an entirely new color scheme, most participants preferred letting the system refresh the palette for them. Even those who made additional adjustments mostly relied on the system-provided swatches for fine-tuning. As U11 noted, ``the system's colors matched the theme and looked good, but I wanted to make slight modifications.''
In the \textit{Layout View}, nine participants browsed alternative layouts and four ultimately chose a non-recommended one, largely for personalize reasons. U1 (\cref{fig:UserStudyGallery}A) switched from the recommended portrait layout to a star layout, noted ``because it looks cool,'' but after further editing in the \textit{Canva View} found that the final arrangement had converged back toward a portrait-like structure similar to InfoAlign's original suggestion.}

\textcolor{black}{Taken together, these interaction patterns show that even when InfoAlign provides a storyframe, visual designs, and spatial composition that are consistent with the story goal, users still actively intervene to adjust  scope, style, and emphasis. \textbf{This highlights that a storytelling infographic must not only align with the story goal but also integrate the creator's subjective intent to form the final communication.} The story-aligned starting point, while users personalize it through small, subjective adjustments, resulting in outputs that remain consistent yet reflect their own intent. Besides, participants described this stepwise involvement as increasing both engagement and trust. As P7 noted, ``Compared with similar tools I've used, InfoAlign's step-by-step creation process feels much more transparent. Being involved at each stage lets me partly `DIY' the infographic, gives me the space to tweak things, and makes me trust the final result more.''}

\subsubsection{Efficiency of Stepwise Human–AI Infographic Authoring}
\textcolor{black}{We examined the efficiency of InfoAlign by analyzing the time spent in each view. On average, participants took 20.3 minutes in total to complete an infographic ($SD = 8.5$). The longest times were observed in the \textit{Story View} and the \textit{Canva View}. In the \textit{Story View} ($M = 6.5$ min, $SD = 3.0$), participants needed to read and interpret the dataset, inspect the generated storyframe and visual designs, and make adjustments, which led to relatively longer engagement. Time spent in the \textit{Stylization View} ($M = 3.3$ min, $SD = 2.4$) and \textit{Layout View} ($M = 2.9$ min, $SD = 1.6$) varied mainly with how much participants wished to creatively fine-tune colors, fonts, and layout. The \textit{Canva View} required the most time ($M = 7.7$ min, $SD = 3.9$), as participants refined the automatically generated blueprint into a final storytelling infographic (see \cref{fig:UserStudyGallery}A, B, and C for examples of the initial generated infographic and the final infographic after modification). Most participants considered this effort acceptable. U11 noted that ``creating one manually would take me much longer, with InfoAlign I only need to make simple adjustments and stylistic tweaks on top of the generated infographic,'' highlighting the efficiency gains of human–AI co-creation. Participants who cared more about visual polish and style (e.g., U1, U5, U10) tended to invest more time, while those primarily concerned with ensuring that the story was correctly and coherently expressed (e.g., U7, U9, U11, U12) finished more quickly. Several participants also suggested that efficiency could be further improved by offering a more flexible and feature-rich canvas, such as Figma (U6).}

\subsection{Overall Effectiveness of InfoAlign}
\label{sec-7.3}
\textcolor{black}{We also evaluated the overall effectiveness of InfoAlign across usability ($M = 6.27$, $SD = 0.84$), creativity support ($M = 6.31$, $SD = 0.75$), human–AI co-creation ($M = 6.10$, $SD = 1.04$), and aesthetics ($M = 6.17$, $SD = 0.88$).} 
In terms of usability, all participants expressed a strong preference for using InfoAlign and agreed that it was easy to learn. Most (11 participants) reported that the system was easy to use and that its functions for storytelling infographic creation were well integrated. As U6 noted, ``I got started quickly after watching the demo video; the operations felt simple, and the system design was very clear.''
All participants affirmed InfoAlign's creativity support across enjoyment, exploration, expressiveness, and perceived value.
Regarding human–AI co-creation, nearly all participants (11) agreed that being able to intervene in the visual designs and layouts helped them better align the output with their desired effect, reinforcing the value of InfoAlign's stepwise intervention design. Whereas fewer participants (9) agreed that they could freely intervene to align the storyframe with their own intended goals. Several participants explained that because the system provides a coherent narrative structure upfront, they sometimes felt steered toward certain organizational patterns, which in turn made it harder to freely re-envision the story according to their own intentions (P6, P8). This reveals a trade-off between automation and expressive freedom, which is discussed further in~\cref{sec-8.1}. 

Almost all participants (11) reported that they sometimes felt they were ``co-creating as partners'' with the system. This suggests that automation provided the scaffolding and user interaction provided the intent: the system guided and recommended, and participants intervened and adjusted the results. From an aesthetic perspective, all participants agreed on the simplicity and clarity of the overall design, as well as the diversity of visual representations. Almost all participants (11) affirmed the colorfulness and craftsmanship of the final outputs. As U4 commented, ``I couldn't have created an infographic of this quality myself, the final result looks very professional.'' Taken together, these results demonstrate InfoAlign's effectiveness in producing consistent, intent-aligned storytelling infographics.

\section{Discussion}
\label{sec-8}
We discuss the implications of our findings and system design, followed by a discussion of current limitations and future directions.

\subsection{Implication}
\label{sec-8.1}
The implications of our findings are as follows.

\textbf{Extending Narrative-Centric Workflows Beyond Infographics.} Although our narrative-centric workflow was originally designed for storytelling infographics, its underlying mechanisms extend naturally to other forms of visual storytelling, such as posters, presentation slides, data comics, and data videos. These formats share similar requirements: constructing a coherent story structure, incorporating visual designs aligned with the story, and organizing spatial composition that preserves narrative flow. User feedback (\cref{sec-7.1}) underscored the value of how the workflow maintains consistent stories throughout the design process. \textcolor{black}{Participants also emphasized InfoAlign's efficiency, its support for creativity, and the aesthetics of the final results, which include visually diverse infographics tailored to different story goals, as illustrated in \cref{fig:Case};} all of these qualities are critical not only for infographics but also for other visual storytelling contexts. Extending narrative-centric workflows to these broader formats could therefore help democratize visual storytelling practices by lowering technical barriers while maintaining narrative consistency. More broadly, such workflows could enhance the accessibility and expressive potential of data-driven storytelling, empowering users across domains to communicate their stories visually.

\textbf{Balancing AI Automation with Human-in-the-Loop Control.} 
\textcolor{black}{Our findings indicate that the AI can automatically generate coherent stories with relatively good efficacy and accuracy (\cref{sec-7.1,sec-7.2}), substantially reducing manual time and labour and, when its suggestions are perceived as reasonable, increasing users' reliance on and trust in the system. However, participants did not simply accept AI-constructed stories as-is: even when the narrative broadly aligned with their goals, they still refined outputs—especially more customized and creative elements such as highlights and icons—while largely keeping core story text and charts as an objective backbone; some also reported that receiving a coherent structure upfront could steer them toward certain organizational patterns (\cref{sec-7.3}), making it harder to freely re-envision the story. These observations suggest that, for design-oriented tools, it is essential to both keep humans in the loop and leave room for user expression.} The core challenge is not whether to automate, but how to balance automation with user control. On the one hand, \textcolor{black}{AI should offer fast, high-quality suggestions that reduce repetitive work and help guide users. On the other hand, users still need enough control to inject their subjective intent and narrative focus so that the final artifact aligns with their design and communication goals. How to balance automation with a sense of authorship remains an important direction for future work.}

\textbf{Opportunities for Richer Input and Adaptive Visual Stories Creation.} 
\textcolor{black}{Although InfoAlign also allows users to convert structured or tabular data into textual input, its narrative-centric workflow is primarily designed for textual input. Tabular data therefore usually serves as local data insight within individual story pieces (e.g., extracted values or simple trends). This reflects a design trade-off: for the same table, different creators may wish to foreground different aspects, so we allow users to convert only the story-relevant parts of a table into text, preserving narrative freedom and designability. However, this also limits the system's ability to fully leverage the structured, quantitative nature of tabular data, constraining the automatic use of richer and more complex quantitative evidence. In addition to tabular data, real-world storytelling often combines multiple information sources, including images, videos, and web content. Different input forms naturally give rise to different narrative structures, which in turn lead to distinct visual story forms. When multiple input types are combined, these narrative structures may interact rather than simply add up. As AI systems move toward richer multimodal input environments, they open opportunities to discover narrative structures tailored to different data forms and to fuse them into richer, more adaptive, and user-specific visual stories.}

\subsection{Limitations and Future Work}
While our work proposes a narrative-centric workflow and introduces InfoAlign as a human–AI co-creation system for storytelling infographic generation, several limitations remain and suggest directions for future enhancement. 
First, InfoAlign provides automated recommendations for visual design and spatial composition, but does not explain the rationale behind these suggestions. For example, although color palettes are aligned with the emotional tone of the story, this reasoning is not explicitly communicated to users. Future iterations could incorporate explanations for AI decisions (e.g., why a certain color is recommended) to improve transparency, strengthen user trust, and support more informed refinements. 
Second, although InfoAlign supports flexible editing, its interaction capabilities are less fine-grained than professional design platforms such as \textit{Figma}\footnote{\url{https://www.figma.com/}} and \textit{Canva}\footnote{\url{https://www.canva.com/}}. Participants noted that advanced editing features, such as localized color adjustments or component-level SVG edits, would enhance creative control. Supporting such refinements while maintaining simplicity is an important future direction.



\section{Conclusion}\label{sec-9}
In this paper, we presented a narrative-centric workflow that maintains story consistency throughout the design of storytelling infographics. The workflow consists of three phases: story construction, story-aligned visual design, and story-informed spatial composition. Building on this workflow, we developed InfoAlign, a human–AI co-creation system that transforms long or unstructured text and user queries into storytelling infographics, while allowing user interventions at each step to preserve story and design intent. \textcolor{black}{We evaluated InfoAlign in a mixed-method user study with 12 infographic creators. The results showed that it consistently supports coherent, intent-aligned storytelling infographics across stages with human–AI co-creation support.} For future work, we plan to extend InfoAlign to support richer multimodal inputs beyond text and to provide adaptive levels of automation that respond to different user expertise and task complexity.

\begin{acks}
This work was supported by Natural Science Foundation of China (NSFC No.62472099).
\end{acks}

\bibliographystyle{ACM-Reference-Format}
\bibliography{main.bib}

\end{document}